%% file: Zhou-TCOM.tex
\documentclass[journal,letterpaper,twocolumn,10pt,twoside]{preamble/IEEEtranTCOM}
\usepackage{hyperref,afterpage,titlesec}
\input{preamble/myLatexCommands}

\usepackage[normalem]{ulem}	%for strikeout only
\usepackage{soul,tcolorbox}
\definecolor{dodgerblue}{rgb}{.06,.3,.55}
\definecolor{forestgreen}{rgb}{.14,.49,.14}
\definecolor{darkblue}{rgb}{.05,.24,.43}

\newcommand{\RNum}[1]{\uppercase\expandafter{\romannumeral #1\relax}}

\begin{document}

%%%%%%%%%%%%%%%%%%%%Journal start%%%%%%%%%%%%%%%%%%%%%%
\title{{Construction D'} Lattices for Power-Constrained Communications}

\author{Fan~Zhou,~\IEEEmembership{Student Member,~IEEE,}
        and Brian~M.~Kurkoski,~\IEEEmembership{Member,~IEEE}% <-this % stops a space
\thanks{This work was supported by JSPS Kakenhi Grant Number JP 19H02137 and 21H04873. This paper was presented in part at the 2021 IEEE International Symposium on Information Theory\cite{Zhou-isit21} and the 2018 International Symposium on Information Theory and Its Applications\cite{zhou-isita18}.}%
\thanks{F.~Zhou and B.~M.~Kurkoski are with the School of Information Science, Japan Advanced Institute of Science and Technology, Nomi 923-1292, Japan (e-mail: fan.zhou \& kurkoski@jaist.ac.jp). \today}
%\thanks{B.~M.~Kurkoski is with the School of Information Science, Japan Advanced Institute of Science and Technology.  F.~Zhou was formerly with School of Information Science, Japan Advanced Institute of Science and Technology and is now with Huawei Technologies.}
}%

%\markboth{IEEE Transactions on Communications}%
%\markboth{ZHOU-TCOM Draft}%
%{Submitted paper}

\maketitle

\begin{abstract}%check the number of words limit: 70~200
%\boldmath
Designs and methods for nested lattice codes using Construction D' lattices for coding and convolutional code lattices for shaping are described. Two encoding methods and a decoding algorithm for {Construction D'} coding lattices that can be used with shaping lattices for power-constrained channels are given. We construct nested lattice codes with good coding properties, a high shaping gain, and low-complexity encoding and decoding. Convolutional code generator polynomials for Construction A lattices with the greatest shaping gain are given, as a result of an extensive search. It is shown that rate $1/3$ convolutional codes provide a more favorable performance-complexity trade-off than rate $1/2$ convolutional codes. Tail-biting convolutional codes have higher shaping gain than that of zero-tailed convolutional codes. A design for quasi-cyclic low-density parity-check (QC-LDPC) codes to form {Construction D'} lattices which have efficient encoding and indexing is presented. The resulting QC-LDPC {Construction D'} lattices are evaluated using four shaping lattices: the $E_8$ lattice, the $BW_{16}$ lattice, the Leech lattice and our best-found convolutional code lattice, showing a shaping gain of approximately $0.65\dB$, $0.86\dB$, $1.03\dB$ and $1.25\dB$ at dimension $2304$.

\end{abstract}


% \fi
\IEEEpeerreviewmaketitle

\section{Introduction} \label{section:introduction}
% for IEEE journal papers produced under \LaTeX\ using
% IEEEtran.cls version 1.7 and later.

\subsection{Motivation}\label{subsection:motivation}

\IEEEPARstart{T}{he} capacity of the additive white {G}aussian noise (AWGN) channel cannot be achieved when equiprobable QAM signal constellations are used\footnote{Probabilistic QAM constellations can provide shaping gain\cite{bocherer-com15}.} at high signal-to-noise ratio (SNR)\cite{Forney-it98}, because they incur a $\pi e/6\ (1.53\dB)$ loss as the dimension $n \rightarrow \infty$. This loss can be overcome using spherical constellations that produce {G}aussian-like distributions, but decoding an $n$-sphere is impractical. Constellation-shaping techniques that produce {G}aussian-like distributions with reasonable complexity are desirable. %Probabilistic shaping performed in the coding scheme proposed by B{\"{o}}cherer, Steiner, and Schulte can result in QAM constellations with shaping gain \cite{bocherer-com15}.  %The two principal classes of codes at high-SNR are trellis codes and lattices. Trellis shaping\cite{Forney-it92} was claimed an asymptotic shaping gain of $1.36\dB$, citing\cite{Marcellin-com90}, using trellis coded modulation\cite{Ungerboeck-it82}.

Lattices are a natural fit for wireless communications because they provide reliable transmission using real-valued algebra and higher transmit power efficiency than conventional constellations at high SNR. Lattices also form an important component of compute-and-forward relaying\cite{Nazer-it11}, which provides high throughput and high spectral efficiency. Voronoi constellations\cite{Conway-it83,Forney-jsac89*2}, also called nested lattice codes\cite{Erez-it04}, constructed using a coding lattice $\Lc$ and a shaping lattice $\Ls$, can be used for power-constrained communications. {E}rez and {Z}amir\cite{Erez-it04} showed that nested lattice codes can achieve the AWGN channel capacity, if $\Lc$ is \emph{channel-good} and the {Voronoi} region of $\Ls$ is hyperspherical, using dithering and minimum mean-square error (MMSE) scaling techniques. For high code rates, dithering is not required\cite{di_pietro-it18} and the role of MMSE scaling becomes negligible\cite{Erez-it04}.

Two lattices $\Lc$ and $\Ls$ are called self-similar if $\Ls$ is an integer-scaled version of $\Lc$. The design of $\Lc$ and $\Ls$ has competing requirements, as $\Lc$ demands good coding properties and an efficient decoding algorithm while $\Ls$ needs good shaping gains and low-complexity quantization. The design of $\Lc$ and $\Ls$ can be separated, under the principle of the separation of coding gain and shaping gain\cite{Forney-it98}. \emph{Rectangular encoding} and \emph{indexing} for non-self-similar nested lattice codes were proposed in\cite{kurkoski-it18}, and conditions on lattice constructions were given.

Low-density parity-check (LDPC) codes have been implemented in a wide variety of communications applications because of their capacity-achievability, efficient encoding, low-complexity decoding, and suitability for hardware implementation. For these reasons, LDPC codes are also suitable for constructing lattices. Lattices based on binary LDPC codes using {Construction D'} were first introduced in\cite{Sadeghi-it06}. Recently Branco~da~Silva and Silva\cite{da_silva-it19} proposed efficient encoding and decoding for {Construction D’} lattices, particularly for LDPC codes. A codeword and cosets of component linear codes are used to form systematic codewords for {Construction D'} lattices. This encoding method naturally produces lattice points in a hypercube. However, a hypercube does not provide shaping gain. A shaping lattice $\Ls$ is needed to do so. 

\subsection{Contributions}\label{subsection:contributions}
In this paper we tackle the encoding and decoding problem of {Construction D'} lattices to be used in power-constrained communications. To achieve shaping gain, convolutional code lattices formed by Construction A are used, with aspects of design, performance and complexity considered. %Both well-known low-dimensional lattices ($E_8$, Barnes-Wall and Leech lattices) as well as convolutional code lattices are used as shaping lattices. 
The main contributions of this paper are as follows. 

We propose two encoding methods and a decoding algorithm for {Construction D'} lattices suitable for power-constrained channels. \emph{Encoding method A} encodes integers with an approximate lower triangular (ALT) check matrix. \emph{Encoding method B} shows how binary information bits are mapped to a lattice point using the check matrix of the underlying nested linear codes for a Construction D’ lattice. Multistage successive cancellation decoding algorithm employing binary decoders is used; the decoder uses re-encoding based on encoding method B; this method is distinct from\cite{da_silva-it19} which was restricted to encoding/decoding Construction D' lattices with hypercubical constellations and cannot achieve shaping gains in a power-constrained channel.
%We present a multistage successive cancellation decoding algorithm employing binary decoders. The \emph{re-encoding} mapping an estimated binary codeword to a lattice point is required during decoding, and this is consistent with encoding method B; these methods are distinct from\cite{da_silva-it19}. 
A definition of {Construction D'} using check-matrix perspective is also given, which is equivalent to the classical congruences definition\cite[p.~235]{Conway-1999}. This is discussed in Section~\ref{section:Dprimelattice}. 

A method to efficiently obtain triangular generator matrices for {Construction A} lattices is given, as a modification of the classical method\cite[p.~183]{Conway-1999} (also \cite[pp.~32--33]{Zamir-2014}). Our method allows forming the generator matrix without swapping code bit positions for convolutional code lattices with underlying rate $1/2,1/3,\ldots$ binary convolutional codes. We show that convolutional code lattices using rate 1/3 codes provide a better tradeoff between shaping gain and quantization complexity than do rate 1/2 codes.  The best-found shaping gains are given in Subsection~\ref{subsection:sgresult}, as a result of an exhaustive search over convolutional code generator polynomials.  %over zero-tailed convolutional codes and tail-biting convolutional codes of rate $1/2$ and $1/3$ with a variety of numbers of states. The asymptotic shaping gains and the complexity of quantization are also given.
%Our results extend the work of\cite{kudryashov-arxiv08,kudryashov-isit10} to a wider range of dimensions, code rates, and additionally considers the shaping gain-complexity tradeoff.} 

%Encoding and indexing\cite{kurkoski-it18}, and construction of nested lattice codes is reviewed in Section~\ref{section:nestedlattice}. 

When applied to high-dimensional nested lattice codes, the conventional indexing algorithm \cite[Sec.~IV-B]{kurkoski-it18} may encounter large-valued integers, which causes an integer overflow when implemented---it may fail to recover information even in the absence of noise. To solve this problem, we modified the algorithm to bound the values of integers that are used internally, without changing the solution; this is described in Subsection~\ref{subsection:encodeindex}. %Hypercube shaping method for {Construction D'} lattices simpler than the conventional hypercube shaping when generating hypercubical constellations is described and implemented for comparison, which is given in Section~\ref{section:simulationresult}.}

We construct quasi-cyclic (QC)-LDPC codes to form {Construction D'} lattices (termed QC-LDPC {Construction D'} lattices), because QC-LDPC codes are widely used in recent wireless communication standards. We give a design of parity-check matrices for nested QC-LDPC codes that can be easily triangularized and thus efficient encoding and indexing is allowed. An existing Construction D' lattice based on a QC-LDPC code and a single parity-check (SPC) product-like code \cite{chen-istc18} is not suitable for indexing/shaping because the SPC product-like code parity-check matrix is cannot be efficiently triangularized. The design and the triangularization method is described in Section~\ref{section:ldpcdesign}. Numerical results of shaping QC-LDPC {Construction D'} lattices using the $E_8$, $BW_{16}$, {Leech} and convolutional code lattice shaping are given in Section~\ref{section:simulationresult}, as well as the comparison with hypercube shaping.

\subsection{Related Work}\label{subsection:relatedwork}
Lattices from linear codes have potential since the decoder for linear codes can be employed to find the nearest lattice point given a point. Well-known methods to build lattices from linear codes are {Construction A} and {D/D'}\cite[Ch.~5, 8]{Conway-1999}. {Construction D/D'} generate lattices from multi-level nested binary linear codes. Binary {Construction A} lattices are the special case of one-level {Construction D} lattices. Unlike {Construction A and D} using generator matrices, {Construction D'} describes lattices by check matrices and thus is suitable for LDPC codes.

Erez and ten Brink employed trellis shaping, constructing lattices based on convolutional codes and {Construction A} that were used for vector quantization in a dirty paper coding scheme\cite{Erez-it05*3}; four rate 1/2 convolutional codes and their shaping gains were given. Kudryashov and Yurkov found generator polynomials of rate $1/2$ convolutional codes that provide the best asymptotic normalized second moment (equivalently, shaping gain) with respect to zero-tailed termination, and near-optimum shaping gain at low dimensions with respect to tail-biting termination in\cite{kudryashov-arxiv08} and\cite{kudryashov-isit10}, respectively. Our results extend their work addressing the optimality of shaping gain to a wider range of dimensions and code rates, and additionally consider the shaping gain-complexity tradeoff. 

%, and our main focus is the shaping gain-quantization complexity tradeoff for a wider selection on rates and dimensions.
%{Construction D} produces lattices with good coding properties and low-complexity decoding by employing the decoder for underlying linear codes. {Construction D} was used to build turbo code lattices\cite{Sakzad-aller10}, polar code lattices\cite{ Yan-itw12} and BCH code lattices\cite{matsumine-glocom18}. 

Using self-similar nested lattice codes, a shaping gain of $0.4\dB$ was shown for low-density lattice codes (LDLCs)\cite{Sommer-itw09}, and a shaping gain of $0.776\dB$ was claimed at $n=60$ for {Construction A} lattices based on QC-LDPC codes\cite{khodaiemehr-com17}. A shaping gain of $0.65\dB$ and $0.86\dB$ was observed using the $E_8$ lattice and the $BW_{16}$ lattice for shaping LDLC lattices, respectively\cite{Ferdinand-twc16}. {Leech} lattice has $1.03\dB$ shaping gain, and was used to shape LDA lattices\cite{di_pietro-com17}. Convolutional code lattices to shape LDLC lattices\cite{Zhou-commlett17} a shaping gain of $0.87\dB$ was preserved at $n=36$. A shaping gain of $0.63\dB$ was found using the $E_8$ lattice for shaping BCH code-based Construction D lattices\cite{buglia-commlett21}.

%Our previous work\cite{Zhou-commlett17} used low-density lattice codes (LDLCs) for coding, which are straightforward to design triangular check matrices. However, it is not simple to obtain a triangular check matrix for an LDPC-based Construction D' lattice. Our previous work\cite{zhou-isita18} studied low-dimensional convolutional code lattices based on zero-tailed convolutional codes. In this paper, moderate-to-high dimensions and tail-biting convolutional codes are included. The work\cite{kurkoski-it18} gave a systematic method for encoding and indexing applicable to non-self-similar nested lattice codes. This indexing algorithm was modified in this Trans.~submission in order to overcome the integer overflow problem at high dimensions.

\subsection{Notation}
A tilde indicates a vector or matrix which has only 0s and 1s --- $\tx$ and $\tH$ are binary while $\x$ and $\H$ are not necessarily so.  Operations over the real numbers $\mathbb R$ are denoted $+,\cdot$ (the operator $\cdot$ is sometimes omitted) while operations over the binary field $\F_2$ are denoted $\oplus, \odot$. The matrix transpose is denoted $\left(\cdot\right)^\textrm{t}$. Element-wise rounding to the nearest integer is denoted $\lfloor \cdot \rceil$.

%\IEEEpubidadjcol

\section{{Construction D'} Lattices} \label{section:Dprimelattice}
We first review the definition of lattices and nested binary codes, then give a definition of {Construction D’} using check-matrix perspective which is equivalent to the congruences definition. Afterwards, how to form lattices from nested binary codes using {Construction D'} is shown. Lastly we propose two equivalent encoding methods and a decoding algorithm for {Construction D'} lattices to be used in power-constrained channels.

\subsection{Preliminaries}

\subsubsection{Lattices}

An $n$-dimensional lattice $\Lambda$ is a discrete additive subgroup of $\Rn$. Let a generator matrix of $\Lambda$ be $\G$ with basis vectors in columns. For integers $\bb \in \Zn$, a vector $\x$ is a lattice point given by $\x=\G\cdot\bb$. We define the \textit{check matrix}\footnote{This is the definition used in low-density lattice codes \cite{Sommer-itw09}, and is distinct from the definition of \cite{Sadeghi-it06}. Note also that the check matrix of a Construction D' lattice is related to, but distinct from, the parity-check matrices of the corresponding binary codes.} as $\H = \G^{-1}$ so that $\H \cdot \x = \bb$. For a lattice with check matrix $\H$, $\x$ is a lattice point if and only if $\H\cdot\x$ is a vector of integers. If $\bW$ is an $n$-by-$n$ unimodular matrix then $\G\cdot\bW$ is also a generator matrix for $\Lambda$.
%Similarly, if $\bW'$ is an $n$-by-$n$ unimodular matrix then $\bW'\cdot\H$ is also a check matrix for $\Lambda$.

\subsubsection{Nested Binary Linear Codes}

\begin{definition}\label{definition:nestedbincodes} Let the row vectors $\h_1,\h_2,\ldots,\h_n$ be a basis for $\F_2^n$. For level $a\geq1$, $\C_0 \subseteq \C_1 \subseteq \cdots \subseteq \C_a = \F_2^n$ are nested linear codes if $\h_{k_i +1},\ldots,\h_n$ are $r_i=n-k_i$ parity-checks for $\C_i$, where $k_i$ denotes the dimension of code $\C_i$ whose rate is $R_i = k_i/n$. That is, a codeword $\tx \in \C_i$ if and only if:
\begin{align}
\h_j \odot \tx = 0,  
\end{align}
for $j = k_i + 1,\ldots,n$ and $i = 0,1,\ldots,a-1$.
\end{definition}

The $n$-by-$n$ matrix of row vectors is denoted 
% \begin{align}
% \tH = [\h_1,\h_2,\ldots,\h_n]
% \label{eqn:matrixtildeH}
% \end{align}%change this to a column if there is enough space
\begin{align}
\tH = 
\renewcommand\arraystretch{0.8}
\begin{bmatrix}
\mbox{---}   \quad   \h_1 \quad \mbox{---}\\
\mbox{---}   \quad   \h_2 \quad \mbox{---}\\
\vphantom{\int h}\smash{\vdots}\\
\mbox{---}   \quad   \h_n \quad \mbox{---}\\
\end{bmatrix}.
\label{eqn:matrixtildeH}
\end{align}

The matrix $\tH_0$ is the parity-check matrix for $\C_0$, and consists of $r_0$ rows, from $\h_{k_0+1}$ to $\h_n$. The matrix $\tH_1$ is the parity-check matrix for $\C_1$, and consists of $r_1$ rows, from $\h_{k_1+1}$ to $\h_n$, and so on. This illustrates that the parity-check matrix for $\C_0$ contains the check matrices for the supercodes $\C_1,\ldots,\C_{a-1}$. The basis vectors $\h_1$ to $\h_{k_0}$ do not contribute to the error-correction capability of the code, but are selected so that $\tH$ is a unimodular matrix, as shown below.

\subsubsection{Construction D'}\label{subsubsection:Dprimedefinitions}

{Construction D'} converts a set of parity-checks defining nested binary linear codes into congruences for a lattice\cite[p.~235]{Conway-1999}. 
A vector %$\x \in \Zn$ 
$\x$ satisfies a congruence $\h = [h_1,\ldots,h_n]$ with respect to a modulo value $q$ if:
\begin{align}
\h\cdot\x\equiv0 \quad (\mod q). \label{eqn:congruences}
\end{align} 
A congruence can be expressed in an equivalent way. Let $\h' = \h/q$. Then $\x$ satisfies this congruence if and only if:
\begin{align} 
\h'\cdot\x \qquad \text{is an integer}. \label{eqn:hxinteger}
\end{align}
Any %$\x \in \Zn$ 
$\x$ satisfying (\ref{eqn:congruences}) will also satisfy (\ref{eqn:hxinteger}). 

Two equivalent definitions of {Construction D'} are given. The classical definition of {Construction D'} uses congruences of parity-checks of nested binary codes.
\begin{definition}[{Construction D'} (congruences)]\cite[p.~235]{Conway-1999}\label{definition:constructionDprimecongruences} 
Let $\C_0 \subseteq \C_1 \subseteq \cdots \subseteq \C_a = \F_2^n$ be nested linear codes. Let the dimension of $\C_i$ be $k_i$. Let $\h_1,\h_2,\ldots,\h_n$ be a basis for $\F_2^n$ such that $\C_i$ is defined by $n-k_i$ parity-check vectors $\h_{k_i+1},\ldots,\h_n$. Then the {Construction D'} lattice is the set of all vectors $\x \in \Zn$ satisfying the congruences:
\begin{align}
\h_j\cdot\x\equiv0 \quad (\mod 2^{i+1}), \label{eqn:conDdefinition1}
\end{align}
for all $i \in \{0,\ldots,a-1\}$ and $k_i+1 \leq j \leq n$.
\end{definition}

Instead of congruences, the following definition uses the check matrix earlier defined as $\mathbf H = \mathbf G^{-1}$.

% \begin{definition}[{Construction D'} (check matrix)] \label{definition:constructionDprime}
% An $n$-by-$n$ unimodular matrix $\tH$ includes the parity-check matrices of nested linear codes $\C_0 \subset \C_1 \subset \cdots \subset \C_a = \F_2^n$. The dimension of $\C_i$ is $k_i$ for $i = 0,1,\ldots,a$, and it has $k_i<k_{i+1}$. Let $\D$ be a diagonal matrix with entries $d_{j,j} = 2^{-i}$ for $k_{i-1} < j \leq k_i$ where $ k_{-1}=0$ and $k_a =n$. Then the {Construction D'} lattice is the set of all vectors $\x$ satisfying: $\H\cdot\x$ are integers, where $\H = \D\cdot\tH$ is the lattice check matrix.
% \end{definition}

\begin{definition}[{Construction D'} (check matrix)] \label{definition:constructionDprime}
An $n$-by-$n$ unimodular matrix $\tH$ includes the parity-check matrices of nested linear codes $\C_0 \subset \C_1 \subset \cdots \subset \C_a = \F_2^n$. The dimension of $\C_i$ is $k_i$ for $i = 0,1,\ldots,a$, and it has $k_i<k_{i+1}$. Let $\D$ be a diagonal matrix with entries:
\begin{align}
d_{j,j} = 2^{-i},
\end{align}
for $k_{i-1} < j \leq k_i$ where $ k_{-1}=0$ and $k_a =n$. Then the {Construction D'} lattice is the set of all vectors $\x$ satisfying:
\begin{align}
\H\cdot\x \qquad \text{are integers,}
\end{align}
where
\begin{align}
\H = \D\cdot\tH \label{eqn:DHtilde}
\end{align}
is the lattice check matrix.
\end{definition}

The following proposition shows that the two definitions are equivalent.
\begin{proposition}
\label{proposition:equivalentDefinitions}
Let $\h_1,\ldots,\h_n$ in Definition \ref{definition:constructionDprimecongruences} be the rows of $\tH$ in Definition \ref{definition:constructionDprime}. Then the lattice given by Definition \ref{definition:constructionDprimecongruences} is identical to the lattice of Definition \ref{definition:constructionDprime}. 
\end{proposition}

\begin{IEEEproof}
It should be clear that because the congruences in (\ref{eqn:conDdefinition1}) can be expressed as (\ref{eqn:hxinteger}), then relevant rows of check matrix $\H$ are an alternative form of the respective congruences. However, our definition of check matrix $\H$ does not include Definition \ref{definition:constructionDprimecongruences}'s restriction to $\x \in \mathbb Z^n$. To acheive this, it is required that $\tH$ be unimodular, so that the {Construction D'} lattice in Definition~\ref{definition:constructionDprime} satisfies $\Lambda \subset \Zn$. To see this, $\G = \H^{-1} = \tH^{-1}\cdot\D^{-1}$. Since $\tH$ is unimodular, $\tH^{-1}$ is an integer matrix. $\D^{-1}$ also is a matrix of integers. Thus $\G$ is an integer matrix and $\Lambda \subset \Zn$. 
\end{IEEEproof}
As a matter of design, after $\tH_0$ to $\tH_{a-1}$ are fixed, the upper rows of $\tH$ should be chosen such that $\tH$ is unimodular; it is also convenient to choose these upper rows so that $\tH$ is ALT form. 

\subsection{Encoding {Construction D'} Lattices} \label{subsection:encDprimelattice}

Two equivalent encoding methods are given. Encoding method A finds a lattice point $\x$ given $\bb\in\Zn$ using its check matrix $\H$ in the ALT form. Encoding method B describes explicitly how information bits $\bu_i$ of the component binary linear code $\C_i$ are mapped to a vector of integers $\bb$ and a lattice point. The two encoding methods can be applied to produce nonhypercubical constellations, which is distinct from the encoding in\cite{da_silva-it19}. 

\subsubsection{Encoding Method A} \label{subsection:encodeA}
Near linear-time encoding of LDPC codes can be accomplished using a parity-check matrix in the ALT form\cite{richardson-it01*3}. This idea inspired us to implement encoding of {Construction D'} lattice $\Lambda$ with a similar procedure. The steps are distinct from\cite{richardson-it01*3} because check matrix $\H$ of $\Lambda$ is a real-valued square matrix.

A vector $\bb$ of integers is provided, which can be considered to be a message sequence. and the corresponding lattice point $\x$ is found by solving: $\H\cdot\x=\bb$.
% \begin{align}
% \H\cdot\x=\bb.  \label{eqn:solveHxeqb}
% \end{align}
If $\H$ is not too big, then $\x$ can be found by matrix inversion: $\x = \H^{-1}\cdot\bb$. If $\H$ is large but is sparse and in the ALT form, as may be expected for {Construction D'} lattices based on LDPC codes, then the following procedure can be used.

Suppose that $\H$ is in the ALT form, that is, it is partially lower triangular. Specifically, $\H$ can be written as:
\begin{align}
\H = 
\renewcommand\arraystretch{0.8}%for double columns
\begin{bmatrix}
\mB &\mA \\
\mX &\mC
\end{bmatrix},\label{eqn:ALTcheckmatrixH}
\end{align}
where $\mA$ is an $s$-by-$s$ lower-triangular matrix with non-zero elements on the diagonal; $\mX$ is a $g$-by-$g$ square matrix. The ``gap'' is $g$---the smaller the gap, the easier the encoding. Let $\Delta = (\mX-\mC\mA^{-1}\mB)^{-1}$. The blockwise inverse\cite{ouellette-laa81} of $\H$ is:
\begin{align}
\H^{-1} = 
\renewcommand\arraystretch{0.8}%for double columns
\begin{bsmallmatrix}
-\Delta\mC\mA^{-1}            &\Delta       \\
\mA^{-1}+\mA^{-1}\mB\Delta\mC\mA^{-1} &-\mA^{-1}\mB\Delta
\end{bsmallmatrix}.
\label{eqn:inverseH}
\end{align}

Using the block structure, $\H\cdot\x=\bb$ can be written as:
\begin{align}
\renewcommand\arraystretch{0.8}
\begin{bmatrix}
\mB &\mA \\
\mX &\mC
\end{bmatrix}
\cdot \renewcommand\arraystretch{0.78}
\begin{bmatrix}
x_1   \\
\vphantom{\int^0}\smash[t]{\vdots}\\
x_g   \\
x_{g+1} \\
\vphantom{\int^0}\smash[t]{\vdots}\\
x_n
\end{bmatrix}
= 
\renewcommand\arraystretch{0.43}
\begin{bmatrix}
b_1   \\
\vphantom{\int^0}\smash{\vdots}\\
b_g   \\
b_{g+1} \\
\vphantom{\int^0}\smash{\vdots}\\
b_n
\end{bmatrix}.
\end{align} %for double columns
% \begin{align}
% %\renewcommand\arraystretch{0.5}
% \begin{bsmallmatrix}
% \mB &\mA \\
% \mX &\mC
% \end{bsmallmatrix}
% \cdot \begin{bsmallmatrix}
% x_1,\ldots,x_g,x_{g+1},\ldots,x_n
% \end{bsmallmatrix}^\textrm{t}
% = \begin{bsmallmatrix}
% b_1,\ldots,b_g,b_{g+1},\ldots,b_n
% \end{bsmallmatrix}^\textrm{t}.
% \end{align}
To perform encoding, first $x_1,\ldots,x_g$ are found using~(\ref{eqn:inverseH}):
\begin{align}
\renewcommand\arraystretch{0.5}
\begin{bmatrix}
x_1   \\
\vphantom{\int^0}\smash[t]{\vdots}\\
x_g   
\end{bmatrix}
=
\begin{bmatrix}
-\Delta\mC\mA^{-1}  &\Delta
\end{bmatrix}
\cdot \bb.
\label{eqn:encodeAstep1}
\end{align}
% \begin{align}
% \begin{bsmallmatrix}
% x_1,\ldots,x_g
% \end{bsmallmatrix}^\textrm{t}
% =
% \begin{bsmallmatrix}
% -\Delta\mC\mA^{-1}  &\Delta
% \end{bsmallmatrix}
% \cdot \bb.
% \label{eqn:encodeAstep1}
% \end{align}
Then, coordinates $x_{g+1},\ldots,x_n$ are found sequentially by back-substitution, using the lower triangular structure of $\H$ which has entry $h_{j,w}$ in row $j$, column $w$. For $w=g+1,\ldots,n$:
\begin{align}
x_w = \frac{1}{h_{j,w}}\bigg(b_j - \sum_{l=1}^{w-1} h_{j,l}x_l\bigg),
\label{eqn:encodeAstep2}
\end{align}
where $j=w-g$.

This method is efficient when $g$ is small and $\H$ is sparse. It uses pre-computation and storage of the $g$-by-$n$ matrix in~(\ref{eqn:encodeAstep1}). The sum in~(\ref{eqn:encodeAstep2}) is performed over few non-zero terms in sparse $\H$. If the check matrix $\H$ is purely triangular, then encoding is simply performed by back-substitution.

\begin{example}\label{eg:encodeA}
Consider a 10-dimensional Construction D' lattice $\Lambda$ generated by nested binary codes $\C_0\subset \C_1$ with parity-check matrix $\tH_0$ and $\tH_1$, respectively. Let $\Lambda$ be described by a check matrix $\H$ in the ALT form, expressed as:
\begin{equation}
\H = 
\begin{tikzpicture}[baseline=0cm,mymatrixenv]
    \matrix [mymatrix,inner sep=2pt,row sep=0.5em,column sep=0.35em] (m)  
    {
        1  &0 &1   &0   &0   &0   &0   &0   &0   &0\\
    0  &0 &1   &1   &0   &0   &0   &0   &0   &0\\
    1  &0 &0   &1   &1   &0   &0   &0   &0   &0\\
    0  &0 &1/2   &1/2   &0   &1/2   &0   &0   &0   &0\\
    1/2 &0  &0   &1/2   &0   &1/2  &1/2   &0   &0   &0\\
    1/2 &0  &1/2  &1/2  &0   &0   &0   &1/2   &0   &0\\
    0  &0 &0   &1/2   &1/2   &0   &0   &1/2   &1/2   &0\\
    0  &0 &0   &1/4   &1/4   &0   &1/4   &1/4   &0   &1/4\\ 
    1/4 &1/4  &1/4   &0   &0   &1/4   &0   &0   &0   &1/4\\
    1/4  &0 &1/4   &1/4   &0   &1/4   &1/4   &0   &1/4   &0\\
    };
    \path (m-1-1.east) -- (m-3-4.west) coordinate[midway] (X)
    (m-8-8.south) -- (m-9-10.north) coordinate[midway] (Y);
    % Dashed lines
    \draw [cheating dash=on 2pt off 2pt,gray]
         (X |- m.north) --   (X |- m.south);
    \draw [cheating dash=on 2pt off 2pt,gray]
         (Y -| m.west) -- (Y -| m.east);

    % Braces
    \hspace{2pt}   
    \mymatrixbraceleft{4}{10}{$\frac{1}{2}\tH_0$}\hspace{8pt}
    \mymatrixbraceleft{8}{10}{$\frac{1}{4}\tH_1$}  
    % \mymatrixbraceleft{4}{10}{$\mathcal{C}_0$}\hspace{25pt}
    % \mymatrixbraceleft{8}{10}{$\mathcal{C}_1$}
\end{tikzpicture}\hspace{8pt},\label{eqn:egencodeAmatrix}
\end{equation}
where the block partition follows (\ref{eqn:ALTcheckmatrixH}).

Assume an arbitrary vector of integers $\bb=[1,2,0,2,4,0,2,0,2,1]^\textrm{t}$. Using (\ref{eqn:encodeAstep1}) the first two positions of the lattice point $\x$ are computed: $x_1=-11, x_2=52$. Then applying (\ref{eqn:encodeAstep2}) we obtain $\x=[-11,52,12,-10,21,2,27,9,-16,-47]^\textrm{t}$.
\end{example}

\subsubsection{Encoding Method B}  \label{subsection:encodeB}
Encoding can also be performed by mapping the message sequence consisting of information vectors $\bu_i \in \F_2^{k_i}$ of $\C_i$ for $i=0,1,\ldots,a-1$ and an integer vector $\z\in \mathbb Z^n$ to a lattice point $\x$. In addition, we show explicitly how $\bu_i$, $\z$ of method B correspond to $\bb$ of method A, with respect to a lattice point $\x$, to establish the equivalence of method A and method B.

For clarity, consider $a = 3$. The integer vector $\bb$ is related to $\bu_0, \bu_1, \bu_2$ and $\z$ as:
\begin{equationarray}{crl}
b_j = &u_{0_j} +2u_{1_j} + 4u_{2_j} + 8z_j,  &\ \text{for $\ 1\leq j\leq k_0,$}\label{eqn:relationbuzbegin}\\
b_j = &         u_{1_j}  + 2u_{2_j} + 4z_j,  &\ \text{for $k_0<j\leq k_1,$} \\
b_j = &                     u_{2_j} + 2z_j,  &\ \text{for $k_1<j\leq k_2,$} \\
b_j = &                                z_j,  &\ \text{for $k_2<j\leq n.$} \label{eqn:relationbuzend}                           \end{equationarray}  
Let $\bu_i'$ be the zero-padded version of $\bu_i$, to have $n$ components:% $\bu_i' = [u_{i_1},u_{i_2},\dots,u_{i_{k_i}},\underbrace{0,\dots,0}_{n-k_i}]^\textrm{t}$. 
\begin{align}
\bu_i' = [u_{i_1},u_{i_2},\dots,u_{i_{k_i}},\underbrace{0,\dots,0}_{n-k_i}]^\textrm{t}.
\end{align}
Then, the integer vector $\bb$ is written as:
\begin{align}
\bb = \D \cdot(\bu_0'+2\bu_1'+4\bu_2'+8\z), \label{eqn:b}
\end{align}
where $\D$ is given in Definition~\ref{definition:constructionDprime}. 

For {Construction D'}, the lattice point $\x$ may be decomposed as: 
\begin{align}
\x = \sum_{i=0}^{a} 2^i \x_i, \label{eqn:xdecomposition}
\end{align}
with components $\x_i$ depending on $\bu_i$ expressed below; $\x_i$ are not necessarily binary. 

Now we describe how information bits are related to a lattice point, and show that recovering integers from a lattice point is possible. Using (\ref{eqn:DHtilde}) and~(\ref{eqn:b})--(\ref{eqn:xdecomposition}) we have
% \begin{align}
%     \H \cdot \x                      &=\bb  \label{eqn:connectxandu}\\
%     \tH \cdot \x                     &=\D^{-1} \cdot \bb \\
%     \tH \cdot(\x_0+2\x_1+\cdots+2^{a-1}\x_{a-1}+2^a\x_a)  &= \bu_0'+2\bu_1'+\cdots+2^{a-1}\bu_{a-1}'+2^a\z, 
% \end{align}
\begin{align}
    \H \cdot \x                      &=\bb,  \label{eqn:connectxandu}\\
    \tH \cdot \x                     &=\D^{-1} \cdot \bb, \\
     \tH \cdot(\x_0+2\x_1+\cdots+2^a\x_a)  &= \bu_0'+2\bu_1'+\cdots+2^a\z,
\end{align}
and the lattice components $\x_i \in \Zn$ satisfy:
\begin{align}
&\tH \cdot \x_i = \bu_i',  \qquad \text{for $i=0,\ldots,a-1$,}   \qquad \text{and} \label{eqn:encstep1} \\
&\tH \cdot \x_a = \z.    \label{eqn:encstep2}
\end{align}
Note that the encoding performed using~(\ref{eqn:encstep1})--(\ref{eqn:encstep2}) is equivalent to encoding method A.

%block diagrams
\begin{figure}[t]
\centering
\includegraphics[width=0.48\textwidth]{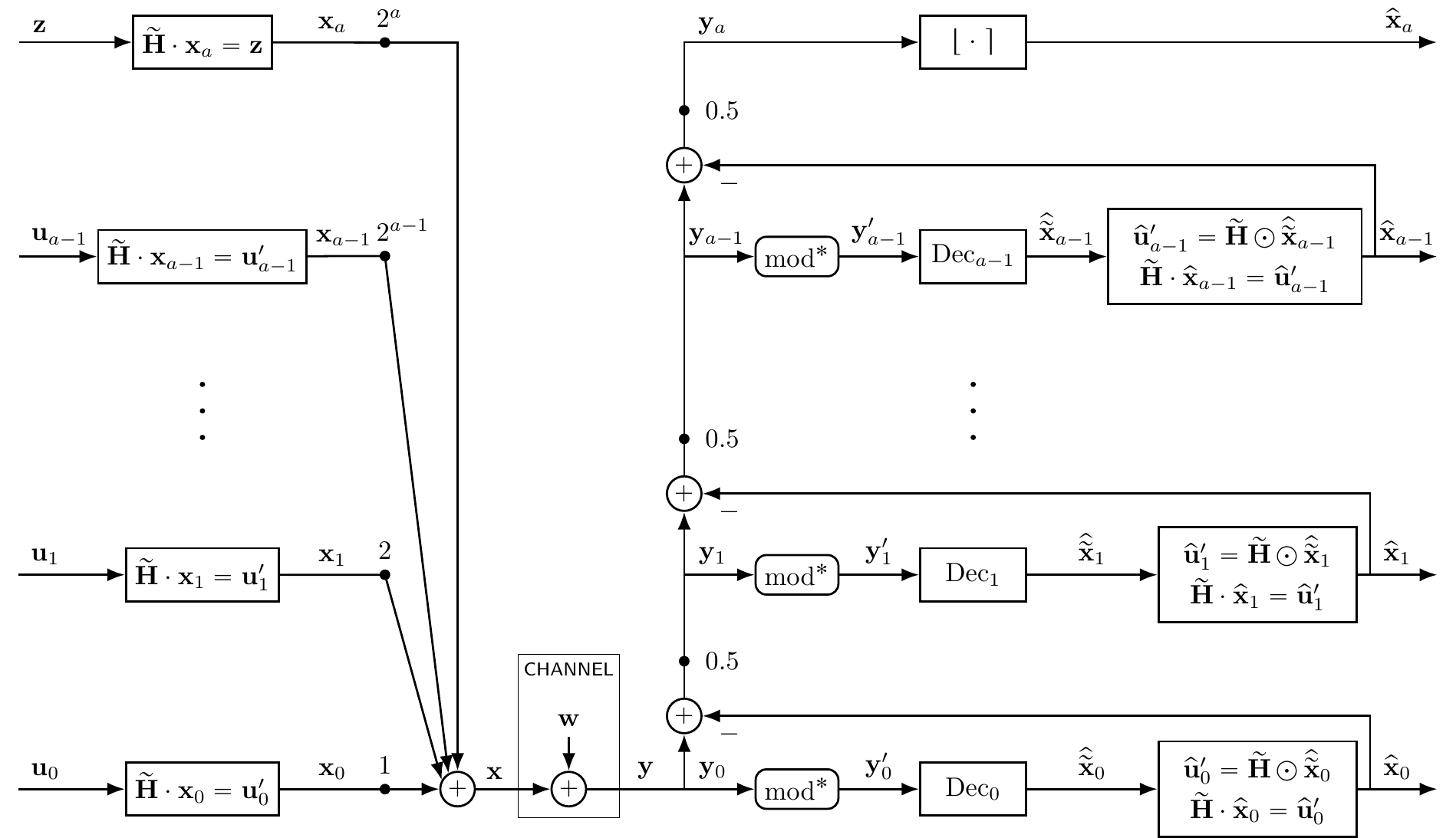}%for double columns
\caption{Block diagram of proposed encoding and decoding {Construction D'} lattices. $\starmod$ denotes the "triangle-function" $\starmod(\y_i) = \left|\modtwo(\y_i+1)-1\right|$ where $\textrm{mod}_2$ indicates a modulo-2 operation.}
\label{fig:blockdiagramDprimelattice}
\end{figure}

\subsection{Decoding {Construction D'} Lattices} \label{subsection:decDprimelattice}

Re-encoding using the generator matrix is typically needed for multistage decoding of Construction D lattices \cite{vem-isit14}. To produce hypercubical constellations with {Construction D'}, multistage decoding may compute cosets instead of re-encoding \cite{da_silva-it19}.  For {Construction D'}, we extend \cite{vem-isit14} and perform re-encoding using the check matrix, and describe a multistage successive cancellation decoding algorithm for Construction D' such that non-hypercubical constellations are allowed. In particular, this decoding algorithm is
suitable for {Construction D'} coding lattices to be used with shaping lattices, likewise employing a binary decoder $\Dec_i$ of $\C_i$, but we use re-encoding corresponding to encoding method B. The encoding and decoding scheme is shown in~Fig.~\ref{fig:blockdiagramDprimelattice}, where encoding method B is to demonstrate the validity of the decoding algorithm.

% Recently\cite{da_silva-it19} multistage decoding that computes the cosets using the estimate of lattice component of all previous levels was proposed for {Construction D'}, \update{producing hypercubical constellations. An earlier paper\cite{vem-isit14} had the idea of using re-encoding when decoding Construction D lattices. We extend re-encoding to check matrix, and describe a multistage successive cancellation decoding algorithm for Construction D' such that non-hypercubical constellations are allowed. In particular, this decoding algorithm is}
% suitable for {Construction D'} coding lattices to be used with shaping lattices, likewise employing a binary decoder $\Dec_i$ of $\C_i$, but we use re-encoding corresponding to encoding method B. The encoding and decoding scheme is shown in~Fig.~\ref{fig:blockdiagramDprimelattice}, where encoding method B is to demonstrate the validity of the decoding algorithm.  

%list the decoding algorithm
\begin{algorithm}[t]\setstretch{1.2} %for double columns
% \begin{figure}[pt]
%   \centering
%   \begin{minipage}{.6\linewidth}
% \begin{algorithm}[H] %for one column
 % \begin{algorithm}%for double column
    \caption{Decoding {Construction D'} Lattices }\label{algorithm:decode}
    \SetKwInOut{Input}{Input}
    \SetKwInOut{Output}{Output}
    \setstretch{1}
\Input{noisy input $\y$}
\Output{estimated lattice point $\hx$}
$\y_0$ = $\y$\;
$\y'_0 = \left|\modtwo(\y_0 +1)-1\right|$\;
$\thx_0 = \Dec_0(\y'_0)$\;
$\hu'_0 = \tH\odot\thx_0$ then solve $\tH\cdot\hx_0 = \hu'_0$\;
\For{ $i = 1,2,\ldots,a-1$}{
  $\y_i = (\y_{i-1}-\hx_{i-1})/2$\;   
  $\y'_i = \left|\modtwo(\y_i +1)-1\right|$\;
  $\thx_i = \Dec_i(\y'_i)$\;
$\hu'_i = \tH\odot\thx_i$ then solve $\tH\cdot\hx_i = \hu'_i$\;
}
$\y_a = (\y_{a-1}-\hx_{a-1})/2$\;
$\hx_a = \lfloor \y_a \rceil$\;
$\hx = \hx_0 + 2\hx_1 + \cdots + 2^{a-1}\hx_{a-1} + 2^{a}\hx_a$
\end{algorithm}
%   \end{minipage}
% \end{figure}

\begin{proposition}\label{proposition}
For {Construction D'}, the lattice component $\x_i$ is congruent modulo 2 to a codeword $\tx_i\in\C_i$, for $i=0,\ldots,a-1$.
\end{proposition}
\begin{IEEEproof}
The lattice component $\x_i$ satisfies $\tH\cdot\x_i=\bu_i'$ and the codeword satisfies $\tH_i\odot\tx_i=\zero$. Recall the last $n-k_i$ positions of $\bu_i'$ are 0s. Row $l$ of $\tH_i$ is equal to row $l+k_i$ of $\tH$, call this row $\h_l$. By definition, $\h_l\cdot\x_i = 0$ and $\h_l\odot\tx_i=0$ for $l=1,2,\ldots,n-k_i$. Thus, $\x_i\, \mod 2=\tx_i$ and the proposition holds.
\end{IEEEproof}

Consider a lattice point $\x$ transmitted over a channel and the received sequence is $\y_0=\x+\w$, where $\w$ is noise. Decoding proceeds recursively for $i=0,1,\ldots,a-1$. The decoding result at level $i-1$ is used before beginning decoding at level $i$. Each level receives $\y_i\in\Rn$ as input, which is mapped to a vector $\y'_i = \left|\modtwo(\y_i +1)-1\right|$ with each element $y'_j\in [0,1]$ for $j=1,2,\ldots,n$. For binary decoders using log-likelihood ratio (LLR) as input, the bit LLR value $\textrm{LLR} = \log\frac{\textrm{Pr}(\widetilde{x}_j=0|y'_j)}{\textrm{Pr}(\widetilde{x}_j=1|y'_j)}$ may be estimated as $\frac{1-2y'_j}{2\var}$. The decoder $\Dec_i$ produces a binary codeword $\thx_i$ closest to $\y'_i$, which is an estimate of $\tx_i$. It is necessary to find $\hx_i$. If $\tx_i$ does not contain an embedded $\hu_i'$, first find $\hu_i' = \tH \odot \thx_i$. Then re-encoding is performed to find $\hx_i$, that is,~(\ref{eqn:encstep1}). This estimated component $\hx_i$ is subtracted from the input, and this is divided over reals by 2: $\y_{i+1} = (\y_i-\hx_i)/2$ to form $\y_{i+1}$, which is passed as input to the next level. This process continues recursively, until $\y_a$ is obtained. The integers are estimated as $\hx_a = \lfloor \y_a \rceil$. The estimated lattice point is written as $\hx = \hx_0+2\hx_1+\cdots+2^a\hx_a$. This successive cancellation decoding is described in Algorithm~\ref{algorithm:decode}.

Furthermore, a {Construction D'} lattice point can also be generated without the need to use the zero-padded $\bu_i$, but is written\footnote{We have $\H\cdot\x = \D\cdot(\tH\cdot\x_0+\cdots+2^{a-1}\tH\cdot\x_{a-1}+2^a\z)$. Recognize that the vector $\tH\cdot\x_i$ is an integer in rows 1 to $k_i$ and is 0 in rows $k_i + 1$ to $n$. The product $2^i\D\cdot\tH\cdot\x_i$ is also an integer vector. Thus, $\H\cdot\x$ is an integer vector. So the decomposition of $\x$ is a lattice point.}  as $\x = 2^a\z+\sum_{i=0}^{a-1} 2^i \x_i$ %for $\z \in \Zn$ 
and the lattice components $\x_i$ should be in a systematic form:
\begin{align}
\x_i = [u_{i_1},u_{i_2},\ldots,u_{i_{k_i}},x_{i_{k_i +1}},\ldots,x_{i_n}]^\textrm{t},    \label{eqn:xsystematicform}
\end{align}
where $x_{i_{k_i +1}},\ldots,x_{i_n}$ are found to satisfy $\tH_i\cdot\x_i=\zero$.
Note that $\x_i$ are not necessarily binary. Therefore decoding {Construction D'} lattices can also be performed using (\ref{eqn:xsystematicform}) for re-encoding; this is  distinct from~Algorithm~\ref{algorithm:decode}. %It is convenient to have binary decoder output an estimate $\hu_i$ of binary information bits $\bu_i$. The decoding proceeds recursively for $i=0,1,\ldots,a-1$ until $\y_a$ is obtained. The integers are estimated as $\hz = \lfloor\y_a\rceil$---this is distinct from~Algorithm~\ref{algorithm:decode} that estimates $\hx_a$ instead of $\hz$. 

%\footnote{This is distinct from choosing $\bb$ and solving $\H\cdot\x=\bb$, which makes the integer-to-bit mapping easy. The trouble comes at the decoder. The component decoders output a binary codeword $\tx$. If $\H\cdot\x=\bb$ is used, this implies a specific encoding mapping. How to recover the original $\bu$ was not clear. Instead, this section aims to obtain $\bu$ from $\bb$---even if some operation is required, it is easier to separate this operation.}:

\section{Nested Lattice Codes} \label{section:nestedlattice}
A coding lattice $\Lc$ and a shaping lattice $\Ls$ are used to form a nested lattice code $\fC$. A practical self-similar $\fC$ in general does not provide both good coding and shaping properties because $\Lc$ and $\Ls$ have competitive design requirements. See\cite{Sommer-itw09,khodaiemehr-com17}. A pair of distinct lattices $\Lc$ and $\Ls$ to form $\fC$ is desirable, where $\Lc$ has good coding properties and $\Ls$ has good shaping properties. This was used in past work: shaping LDLC lattices using the $E_8$ lattice and the $BW_{16}$ lattice\cite{Ferdinand-twc16}, convolutional code lattices \cite{Zhou-commlett17}, and shaping LDA lattices using the {Leech} lattice \cite{di_pietro-com17}. These results show that $\Lc$ and $\Ls$ can be designed to provide both good coding properties and efficiently achievable shaping gains.

Encoding (mapping information to lattice codewords) and indexing (the inverse mapping) of nested lattice codes can be performed if the check matrix of $\Lc$ and the generator matrix of $\Ls$ are triangularizable. The matrices used for encoding and indexing are assumed lower triangular. The encoding method follows\cite{kurkoski-it18} is briefly reviewed. Then we modify the indexing method in\cite[Sec.~IV-B]{kurkoski-it18} so that bounding values for integers are found, thus overcoming the integer overflow problem for high-dimensional lattices. After that, the coding scheme used in this paper is described.

\subsection{Encoding and Indexing of Nested Lattice Codes} \label{subsection:encodeindex}
%In this subsection the condition to form a nested lattice code is reviewed. The encoding method follows\cite{kurkoski-it18} thus is described briefly. Then we propose an indexing method that overcomes the integer overflow problem.

\subsubsection{Preliminaries}
Let $\Gc$ and $\Gs$ be a generator matrix of $\Lc$ and $\Ls$ respectively. The check matrix of $\Lc$ is $\Hc = \Gc^{-1}$. To build a nested lattice code, a coding lattice $\Lc$ and a shaping lattice $\Ls$ satisfy $\Ls \subseteq \Lc$ (is referred to as the \emph{sublattice condition}\cite[p.~179]{Zamir-2014}).% 2) $\Lc/\Ls$ forms a quotient group. 
%Recall the following lemma for the sublattice condition. 
\begin{lemma}\cite[Lemma 1]{kurkoski-it18}
$\Ls\subseteq\Lc$ if and only if $\Hc\Gs$ is a matrix of integers.
\label{lemma:sublattice}
\end{lemma}
% \begin{IEEEproof}
% Let $\Gs\bb\in\Ls$. The point $\Gs\bb$ is a point in $\Lc$ if and only if $\Hc\Gs\bb$ is a vector of integers. For arbitrary $\bb\in\Zn$, this is true if and only if $\Hc\Gs$ is a matrix of integers.
% \end{IEEEproof}

%Let the fundamental region be the zero-centered {Voronoi} region $\V$ of $\Ls$, then the set of coset leaders of a quotient group $\Lc/\Ls$ is the codebook, or lattice code $\fC = \Lc \cap \V$.
A nested lattice code can be defined by $\fC= \Lc \cap \V$ where $\V$ is the zero-centered {Voronoi} region of the shaping lattice $\Ls$. The codebook is the set of coset leaders of a quotient group $\Lc/\Ls$.

\subsubsection{Encoding} \label{subsection:encoding}
The mapping from integers to a lattice codeword in $\fC$ is called encoding \cite{kurkoski-it18}. Assume that $\Hc$ and $\Gs$ are lower triangular. Let $h_{i,i}$ and $g_{i,i}$ be diagonal elements of $\Hc$ and $\Gs$ for $i=1,2,\ldots,n$. It follows that $M_i = h_{i,i}g_{i,i}$ is a positive integer. Let information be represented by a vector of integers $\bb$ where $b_i \in \{0,1,\ldots,M_i -1\}$ and position $i$ encodes $\log_2 M_i$ bits. Encoding is bijectively mapping $\bb$ to $\x'\in\fC$, where the number of codewords is $|\fC|=\prod_{i=1}^n M_i$. The lattice codeword is given by:
% \begin{align}
% \x' = \Gc\bb \,\mod\Ls.    \label{eqn:latticemod}
% \end{align}
\begin{align}
\x' = \x \,\mod\Ls,    \label{eqn:latticemod}
\end{align}
where $\x\in\Lc$ can be found by solving $\Hc\x=\bb$ using the methods in Section~\ref{section:Dprimelattice}. Here $\Hc$ need not be lower triangular but needs to be triangularizable using a unimodular transformation. Note that dithering is omitted when we discuss encoding and indexing for simplicity, and will be described in Subsection~\ref{subsection:codingscheme}.

\begin{example}\label{eg:encnestedlatticecode}
Let the shaping lattice $\Ls=4\Lambda_{A}^{10}$ be described by a generator matrix $\Gs$, which is the scaled-by-4 version of the matrix $\GLA^{10}$ that will be given in Example~\ref{eg:ccltri} in Section~\ref{section:convshaping}. Let $\Hc$ be the triangularized version of the check matrix $\H$ in equation (\ref{eqn:egencodeAmatrix}) with $\bW\cdot\H=\Hc$ for a unimodular matrix $\bW$. The lower-triangular matrix
\begin{align}
\Hc=\renewcommand{\arraystretch}{1}
    \begin{bsmallmatrix}
1        & 0         &0      &   0      &   0       &  0       &  0       &  0      &   0     &    0\\
0       &  1        & 0     &    0     &    0      &   0     &    0    &     0     &    0        & 0\\
0      &   0       &  1     &    0      &   0     &    0      &   0      &   0       &  0    &     0\\
0      &-1/2      &   0     &  1/2        & 0      &   0    &     0     &    0      &   0  &       0\\
-1/2    &  -1/2      &   0      & 1/2      & 1/2    &     0     &    0      &   0     &    0      &   0\\
0     &    0    &   1/2     &  1/2       &  0     &  1/2        & 0       &  0       &  0      &   0\\
1/2      &   0       &  0      & 1/2      &   0     &  1/2      & 1/2       &  0       &  0      &   0\\
-1/4    &  -1/4     & -1/4      & 1/4     &  1/4    &  -1/4     &  1/4     &  1/4      &   0     &    0\\
1/4      &   0     &  1/4     &  1/4      &   0     &  1/4     &  1/4      &   0      & 1/4      &   0\\   
1/4      & 1/4      & 1/4      &   0     &    0     &  1/4       &  0       &  0       &  0     &  1/4
    \end{bsmallmatrix}
\end{align}
% \begin{align}
%     \bW=\renewcommand{\arraystretch}{0.55}
%     \begin{bmatrix}
%     -1 & -1 & 1 & 0 & -2 & 2 & -2 & 0 & 0 & 4 \\ 
% 4 & 4 & -3 & -4 & 10 & -6 & 8 & -4 & 4 & -16 \\
% 2 & 1 & -1 & 0 & 2 & -2 & 2 & 0 & 0 & -4 \\    
% -3 & -2 & 2 & 2 & -6 & 4 & -5 & 2 & -2 & 10 \\ 
% -1 & -1 & 1 & 2 & -3 & 1 & -2 & 2 & -2 & 4 \\  
% 0 & 0 & 0 & 1 & 0 & 0 & 0 & 0 & 0 & 0 \\       
% 0 & 0 & 0 & 0 & 1 & 0 & 0 & 0 & 0 & 0 \\       
% 0 & 0 & 0 & 0 & 0 & 0 & 0 & 1 & -1 & 0 \\      
% 0 & 0 & 0 & 0 & 0 & 0 & 0 & 0 & 0 & 1 \\       
% 0 & 0 & 0 & 0 & 0 & 0 & 0 & 0 & 1 & 0 
%     \end{bmatrix}
% \end{align}
%\footnote{\update{The integer vector in Example~\ref{eg:encodeA} does produce a lattice point of $\Lc$ but cannot be used to generate a codeword of the nested lattice code because there exist integers not in $\{0,1,\ldots,M_i -1\}$.}}
is used when encoding and indexing. The diagonal elements $M_i$ of $\Hc\Gs$ for $i=1,2,\ldots,10$ are: $M_i\in\{4,8,4,4,2,4,4,2,2,2\}$, % \begin{align}
%     M_i\in\{4,8,4,4,2,4,4,2,2,2\},
% \end{align}
which gives the range of information integers. Then the code rate is $R=\frac{1}{10}\,\log_2 \prod_{i=1}^{10} M_i=1.7$ bits per dimension. Assume the information vector\footnote{The corresponding information bits are $\bu_0=[0,0,1]^\textrm{t}$ and $\bu_1=[1,0,0,0,0,0,0]^\textrm{t}$ for the underlying binary codes $\C_0$ and $\C_1$ of $\Lc$, respectively. The remaining information bit positions in $\bb$ may be selected using integers $\z$ similar to (\ref{eqn:relationbuzbegin})--(\ref{eqn:relationbuzend}). Under correct decoding, these $\bu_0, \bu_1$ and $\z$ are produced by each level of the decoder. Note that the matrices used for encoding and the decoder's re-encoding should agree.} is: $\bb=[2,4,1,2,0,0,2,1,0,0]^\textrm{t}$. By solving $\Hc\x=\bb$ using back-substitution a lattice point $\x=[2,4,1,8,-2,-9,3,-7,-5,2]^\textrm{t}$ is generated. The shaping operation (\ref{eqn:latticemod}) or equivalently (\ref{eqn:latticeencode}) using $4\Lambda_{A}^{10}$ gives $\x'=[-2,0,-3,0,-2,-1,-1,1,-1,-2]^\textrm{t}$. In this example, $4\Lambda_{A}^{10}$ has a shaping gain of $0.58\dB$ which is obtainable because any lattice codeword $\x'$ lies in the zero-centered Voronoi region of $4\Lambda_{A}^{10}$---this produces a nonhypercubical constellation.% \update{Increasing the dimension, the shaping gain of convolutional code lattices as will be given in Fig.~\ref{fig:cclsg} can be obtained.}
\end{example}

\subsubsection{Indexing}\label{subsection:indexing}
The inverse of encoding is called indexing that maps a lattice codeword $\x' \in \fC$ to the vector of integers $\bb$ used by the encoder. %in the {Voronoi} region of $\Ls$ to a point of $\Lc$ inside the parallelotope. 
Note that $\x'$ and $\x=\Gc\bb$ are in the same coset, so when $\x \neq \x'$, in general, using $\Hc\x'$ cannot recover $\bb$ and thus an indexing method is necessary. This can be done by a systematic procedure as suggested in\cite[Sec.~IV-B]{kurkoski-it18}. The modulo-$\Ls$ expression~(\ref{eqn:latticemod}) can also be written as 
\begin{align}
    \x'=\Gc\bb-Q_{\Ls}(\Gc\bb), \label{eqn:latticeencode}
\end{align}
where $Q_{\Ls}$ is a lattice quantizer that finds the nearest lattice point in $\Ls$ given a point. Let $\bb' = \Hc\x'$. Multiply $\Hc$ on the left of both sides of (\ref{eqn:latticeencode}) so that $\bb' = \bb-\Hc Q_{\Ls}(\Gc\bb)$. The indexing can be performed by finding $\bt\in\Zn$ that satisfies $Q_{\Ls}(\Gc\bb) = \Gs\bt$ such that
\begin{align}
\bb' = \bb-\Hc\Gs\bt. 
\label{eqn:latticeindexold}
\end{align}
The indexing algorithm was described in\cite[Sec.~IV-B]{kurkoski-it18}. Consider high-dimensional nested lattice codes. %Observe that the solution to\cite[eq.~(39)]{kurkoski-it18} can easily 
As the integers $b_i, t_i$ are found sequentially, the values for $t_i$ can become large which leads to an integer overflow problem in practical implementations, depending on the elements of $\Hc\Gs$ and especially when $\Gs$ has large scaling.

Now we propose a modified method suitable for indexing high-dimensional nested lattice codes. Instead of using (\ref{eqn:latticeindexold}) we introduce $\s\in\Zn$ such that
\begin{align}
    \bb' = \bb+\Hc\Gs\s-\Hc\Gs\e,    \label{eqn:latticeindex}
\end{align}
where $\e=\bt+\s$ will be shown to be bounded. The solution $\bb$ can be found without explicitly computing $\bt$ and $\s$, thus the integer overflow problem can be avoided. 

These equations are solved sequentially first for $i = 1$, then $i = 2,\ldots,n$, using the triangular structure of $\Hc\Gs$ which is expressed as
\begin{align}
   \Hc\Gs = \renewcommand\arraystretch{0.8}
   \begin{bmatrix}
   \theta_{1,1} &0               &\cdots         &0 \\
   \theta_{2,1} &\theta_{2,2}    &\cdots         &0 \\
  \vphantom{\int }\smash[^t]{\vdots}       &\vphantom{\int }\smash[^t]{\vdots}          &\vphantom{\int}\smash[^t]{\ddots}         &\vphantom{\int}\smash[^t]{\vdots} \\
   \theta_{n,1} &\theta_{n,2}    &\cdots       &\theta_{n,n}
   \end{bmatrix},
\end{align}
where $\theta_{i,i}=M_i$. The first line of (\ref{eqn:latticeindex}) is
\begin{align}
    b'_1 = b_1 + M_1 s_1 - M_1 e_1. \label{eqn:indexlineone1}
\end{align} 
Then for $i = 2,\ldots,n$:
\begin{align}
    b'_i = b_i +\sum_{j=1}^{i-1} \theta_{i,j}s_j + M_i s_i - \sum_{j=1}^{i-1} \theta_{i,j}e_j - M_i e_i. \label{eqn:indexbtildeoffset}
\end{align}
Firstly, the solution of $b_i$ is found as follows. To obtain $e_i$ we write
\begin{align}
q_i=s_i+\sum_{j=1}^{i-1} \frac{\theta_{i,j}}{M_i}s_j,
\end{align}
%$q_i=s_i+\sum_{j=1}^{i-1} \frac{\theta_{i,j}}{M_i}s_j$ 
but $s_i$ need not to be computed. Then $q_i$ should be chosen such that $e_i$ is bounded and after $e_i$ is obtained as indexing proceeds, the value is used for $i+1,\ldots,n$. The solution $\bb$ of (\ref{eqn:latticeindex}) is the same as that of (\ref{eqn:latticeindexold}) by choosing $q_i$ such that
\begin{align}
q_i/\lcm(M_{i+1},\ldots,M_n) \qquad \text{is an integer.}
\end{align}
% $q_i/\lcm(M_{i+1},\ldots,M_n)$ is an integer. %The solution $\bb$ of (\ref{eqn:latticeindex}) is the same as that of (\ref{eqn:latticeindexold}) if $\lcm(M_{i+1},\ldots,M_n)|s_i$ and $\lcm(M_{i+1},\ldots,M_n)|q_i$. 

The algorithm is given as follows. The solution of (\ref{eqn:indexlineone1}) is $b_1$ and $e_1$ given by
\begin{align}
    b_1 &=b'_1 \,\mod M_1, \qquad\text{and}\\
    e_1 &= \frac{b_1-b'_1}{M_1} \,\mod \lcm(M_2,M_3,\ldots,M_n).
\end{align}
%where $q_1=\delta\lcm(M_2,M_3,\ldots,M_n)$ for $\delta\in\Z$. The integer $\delta$ is chosen such that $e_1$ is bounded.

Then for $i = 2,\ldots,n$, (\ref{eqn:indexbtildeoffset}) has solution $b_i$ and $e_i$ given by
\begin{align}
    b_i &= b'_i + \sum_{j=1}^{i-1} \theta_{i,j}e_j \,\mod M_i, \qquad\text{and} \label{eqn:indexsolveb} \\
    e_i &= \frac{b_i-b'_i - \sum_{j=1}^{i-1} \theta_{i,j}e_j}{M_i} \,\mod \lcm(M_{i+1},\ldots,M_n), \label{eqn:indexsolvee}
\end{align}
where the integer $0\leq e_i< \lcm(M_{i+1},\ldots,M_n)$ is thus bounded---this is practical.

Triangular $\Hc$ and $\Gs$ allow efficient encoding and indexing, where $\Hc$ and $\Gs$ can be obtained from triangularizable full-rank check matrix and generator matrix of $\Lc$ and $\Ls$ respectively. We have not yet found a straightforward method to index nested lattice codes using non-triangular matrices. 

\subsection{Coding Scheme}\label{subsection:codingscheme}

Erez and Zamir\cite{Erez-it04} proposed a coding scheme using nested lattice codes with dithering and MMSE scaling techniques that can achieve the capacity of the power-constrained AWGN channel, which is transformed into a modulo-lattice additive noise channel. %We use a coding scheme similar to\cite{Erez-it04,Zhou-commlett17} and is given in~Fig.~\ref{fig:blockdiagramnestedlattice}. But\cite{Erez-it04} did not consider indexing, and\cite{Zhou-commlett17} did not use dithering and MMSE scaling. 
We use a similar coding scheme, but additionally include the indexing. Since this paper considers primarily high rate codes in the high-SNR domain, the MMSE scaling is close to 1. As proven by di~Pietro, Z{\'e}mor, and Boutros\cite{di_pietro-it18}, dithering is not mandatory because lattice points of $\Lc$ at high code rate fill well the Voronoi region of $\Ls$.

Let the dither $\bU$ be uniformly distributed in {Voronoi} region of $\Ls$, which is independent of the lattice point $\x$ of $\Lc$. Instead of using (\ref{eqn:latticemod}), a vector ${\x''}=\x-\bU \;\textrm{mod}\;\Ls$ is sent to the AWGN channel. The average transmitted power per symbol $\es=\frac{1}{n}\mathrm{E}[\Vert{\x''}\Vert^2] = \frac{1}{n}\mathrm{E}[\Vert\bU\Vert^2]$ can also be represented by $\es=\textrm{NSM}\cdot V^{2/n}(\Ls)$ where $\Ls$ has normalized second moment NSM\cite[eq.~(1)]{Conway-siam84} and volume $V(\Ls)$. The MMSE scaling coefficient $\alpha$ is defined $\alpha = \es/(\es+\var)$ where $0 \leq \alpha \leq 1$.%$\alpha = \frac{\es}{\es+\var} = \frac{\SNR}{1+\SNR}$ where $0 \leq \alpha \leq 1$.
The signal-to-noise ratio is defined as $\SNR = \es/\var$. Thus $\alpha$ can also be expressed $\alpha = \SNR/(1+\SNR)$. Given a received sequence ${\y''}={\x''}+\w$ where $\w$ is noise, the input to the decoder is computed $\y=\alpha{\y''}+\bU$. See\cite{Erez-it04,di_pietro-com17}.

The rate of a nested lattice code $\fC$ is defined: %$R = \frac{1}{n}\log_2 \vert\fC\vert = \frac{1}{n}\log_2\frac{\vert\det(\Gs)\vert}{\vert\det(\Gc)\vert}$.
\begin{align}
R = \frac{1}{n}\log_2 \big|\fC\big| = \frac{1}{n}\log_2\frac{\big|\det(\Gs)\big|}{\big|\det(\Gc)\big|}.
\end{align}
The average transmitted power per bit can be computed $\eb = \es/R$. In this paper we measure the decoding error rate of nested lattice codes as a function of $\ebnoshort = \eb/2\var = \SNR/2R$. To observe the shaping gains, it is convenient to define the {Shannon} limit in terms of $\ebnoshort$ as $10\log_{10}(2^{2R}-1)/(2R)$ given in decibels.

\begin{figure}[pt]
\centering
\includegraphics[width=0.48\textwidth]{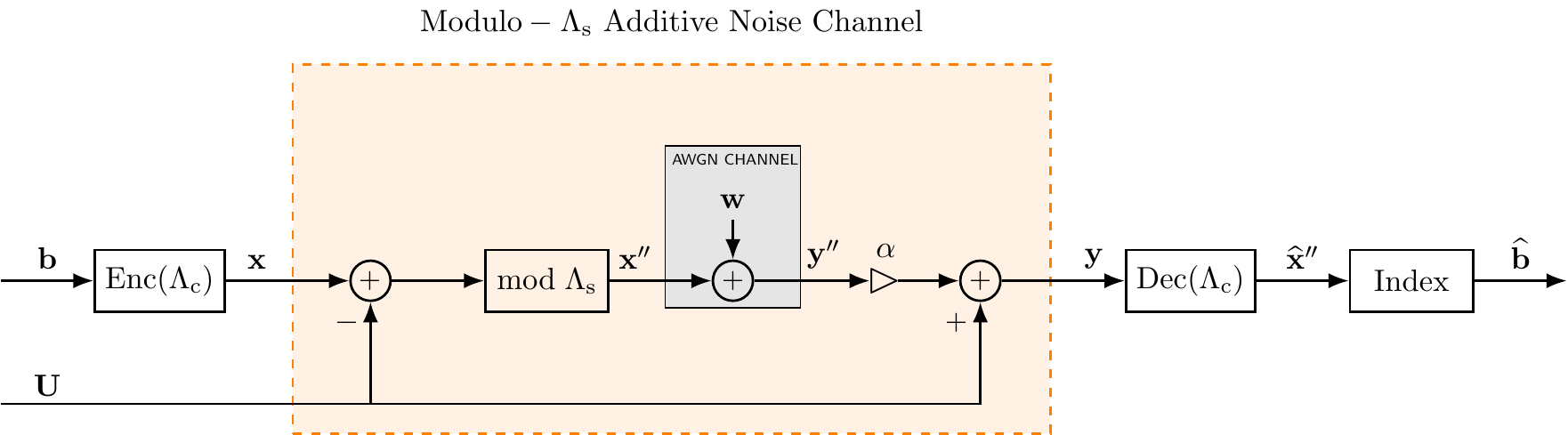}%for double columns
\caption{Block diagram of nested lattice codes with a dither variable $\bU$ uniformly distributed over the {Voronoi} region of $\Ls$ and a ``Wiener coefficient'' $\alpha$ was chosen for MMSE.}
\label{fig:blockdiagramnestedlattice}
\end{figure}

\section{Convolutional Code Lattices} \label{section:convshaping}
This section focuses on the design of convolutional code lattices which are Construction A lattices using convolutional codes. The zero-centered Voronoi region $\V$ of a convolutional code lattice is used to construct the nested lattice code. The effectiveness of an $n$-dimensional lattice quantizer is measured by the shaping gain with respect to the normalized second moment of $\V$ and that of the integer lattice $\Zn$. The shaping gain measures the signal power reduction, and the theoretic limit is $1.53\dB$ given by an $n$-sphere as $n\rightarrow\infty$\cite{Forney-it98}.

%Lattices built from convolutional codes using {Construction~A} (we refer to them as convolutional code lattices) were used for vector quantization and the shaping gain of four lattices were provided\cite{Erez-it05*3}. 
The shaping gains of convolutional code lattices were studied in\cite{Erez-it05*3,kudryashov-arxiv08,kudryashov-isit10,Zhou-commlett17,zhou-isita18}. Convolutional code lattices have high shaping gain, flexibility of lattice dimension, and low-complexity quantization using the well-known {Viterbi} algorithm. For these reasons, convolutional code lattices are suitable as shaping lattices. We are interested in both shaping gain and the complexity of shaping. %Convolutional code lattices are constructed using convolutional codes and {Construction A}\cite{Conway-1999}. 

We give a method to obtain triangular generator matrices for {Construction A} lattices that is modified from\cite{Conway-1999,Zamir-2014,Zhou-commlett17,zhou-isita18}. This is applied to build convolutional code lattices from zero-tailed convolutional codes and tail-biting convolutional codes. An exhaustive search finding the convolutional code generator polynomial that provides best-found shaping gain is performed. The tradeoff between shaping gain and quantization complexity of both zero-tailed convolutional codes and tail-biting convolutional codes is also studied.

\subsection{Triangular Matrix of {Construction A} Lattices}\label{subsection:trimatrixA}
{Construction A} with binary codes corresponds to the case of one-level {Construction D}. Triangular matrices provide convenient encoding and indexing, thus we discuss how to obtain a triangular generator matrix $\GLA$ for a {Construction A} lattice $\LA$. The well-known methods in\cite[p.~183]{Conway-1999} and \cite[pp.~32--33]{Zamir-2014} require a systematic generator matrix for the code. The method given below does not require a systematic code generator matrix; while convolutional codes do have a systematic form it requires swapping bit positions (or coordinate permutation). Also, our method produces matrices already in the Hermite normal form as defined in\cite[pp.~42--44]{costa-2017} for forming Construction A generator matrices. 

Let $\G'=[\g_1',\g_2',\ldots,\g_k']$ be an $n$-by-$k$ full-rank generator matrix with basis vectors in columns for a binary code $\C$. Perform column operations on $\G'$ to find $\G=[\g_1,\g_2,\ldots,\g_k]$ where $\G$ has the property that for each column $i = 1, \ldots, k$, there are only zeros to the right of the first one in column $i$. The canonical form for rate $1/2,1/3,\ldots$ zero-tailed convolutional codes already satisfy this condition. Let $\I_n$ be an $n$-by-$n$ identity matrix. The lower-triangular generator matrix $\GLA$ of a {Construction A} lattice $\LA$ can be obtained by replacing $k$ columns in $2\I_n$ using the columns in $\G$. If $\g_i$ has its first one in position $j$, then replace column $j$ of $2\I_n$ with $\g_i$, for all $i$. As a Construction A lattice, the determinant is $\det(\LA)=\det(\GLA) = 2^{n-k}$.

\begin{example}
\label{eg:ccltri}
Consider a generator matrix $\G'$ of a nonsystematic feedforward zero-tailed convolutional code with generator polynomials represented as octal numbers $[7,5]$, where the information sequence has length 3. Then apply Construction A to form a lattice $\Lambda_{A}^{10}$ by replacing the 3 columns in $2\I_{10}$ using the columns in $\G'$, resulting in a lower-triangular generator matrix $\GLA^{10}$. This is expressed as:
\begin{align}
\G'= 
\renewcommand\arraystretch{0.45}
\begin{bsmallmatrix}
     1     &0     &0\\
     1     &0     &0\\
     1     &1     &0\\
     0     &1     &0\\
     1     &1     &1\\
     1     &0     &1\\
     0     &1     &1\\
     0     &1     &0\\
     0     &0     &1\\
     0     &0     &1
\end{bsmallmatrix}\quad \Longrightarrow \quad
\GLA^{10}= 
\renewcommand\arraystretch{0.45}
\begin{bsmallmatrix}
     1     &0     &0     &0     &0     &0     &0     &0     &0     &0\\
     1     &2     &0     &0     &0     &0     &0     &0     &0     &0\\
     1     &0     &1     &0     &0     &0     &0     &0     &0     &0\\
     0     &0     &1     &2     &0     &0     &0     &0     &0     &0\\
     1     &0     &1     &0     &1     &0     &0     &0     &0     &0\\
     1     &0     &0     &0     &1     &2     &0     &0     &0     &0\\
     0     &0     &1     &0     &1     &0     &2     &0     &0     &0\\
     0     &0     &1     &0     &0     &0     &0     &2     &0     &0\\
     0     &0     &0     &0     &1     &0     &0     &0     &2     &0\\
     0     &0     &0     &0     &1     &0     &0     &0     &0     &2
\end{bsmallmatrix}.
\end{align}
This convolutional code lattice has a shaping gain of $0.58\dB$, obtained using the Viterbi algorithm for quantization.
\end{example}
%example about arbitrary binary linear code
% \begin{example}
% Consider a full-rank generator matrix $\G'$ of an arbitrary binary linear code $\C$. Replace the second column of $\G'$ by the sum of the first two columns to obtain $\G$:
% \begin{align}
% \G'= 
% %\renewcommand\arraystretch{0.9} %for double columns
% \renewcommand\arraystretch{0.5}
% \begin{bmatrix}
%      1     &1     &0\\
%      1     &1     &0\\
%      1     &0     &0\\
%      0     &1     &0\\
%      1     &0     &0\\
%      1     &1     &1
% \end{bmatrix}\quad \Longrightarrow \quad
% \G= 
% %\renewcommand\arraystretch{0.9} %for double columns
% \renewcommand\arraystretch{0.5}
% \begin{bmatrix}
%      1    &0    &0\\
%      1    &0    &0\\
%      1    &1    &0\\
%      0    &1    &0\\
%      1    &1    &0\\
%      1    &0    &1
% \end{bmatrix}
% \end{align}
% Apply {Construction A} to form a lattice $\LA$ by adding 3 columns to $\G$, then the generator matrix $\GLA$ is given by:
% \begin{align}
% \GLA= 
% %\renewcommand\arraystretch{0.9} %for double columns
% \renewcommand\arraystretch{0.5}
% \begin{bmatrix}
%      1     &0    &0     &0     &0     &0\\
%      1     &2    &0     &0     &0     &0\\
%      1     &0    &1     &0     &0     &0\\
%      0     &0    &1     &2     &0     &0\\
%      1     &0    &1     &0     &2     &0\\
%      1     &0    &0     &0     &0     &1
% \end{bmatrix}.
% \end{align}
% \end{example}

% The code in the example is not a convolutional code, but was chosen to illustrate the construction of $\GLA$. As a Construction A lattice, the determinant is $\det(\LA)=\det(\GLA) = 2^{n-k}$.

\subsection{Best-Found Convolutional Code Lattices} \label{subsection:sgresult}

In this subsection, rate $1/2$ and $1/3$ convolutional codes with nonsystematic feedforward encoders are used to build {Construction A} lattices $\LA$. Let $m$ be the memory order of convolutional code encoders. The number of states is $2^m$.

The generator matrix of zero-tailed convolutional codes has the desired form described in the previous subsection, and thus is straightforward to find a lower-triangular generator matrix for convolutional code lattices. Let $R_{\textrm{ZTCC}}$ be the code rate of a zero-tailed convolutional code. The information length is $k=nR_{\textrm{ZTCC}}-m$. Therefore rate loss exists and it affects the code performance when $n$ is small.

\begin{figure}[!t]
\centering
\includegraphics[width=0.48\textwidth]{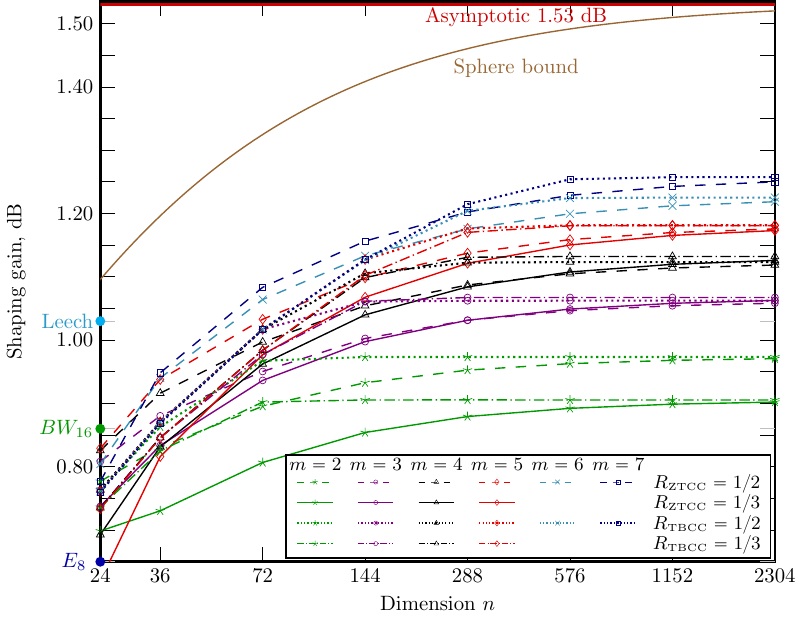} %for double columns
\caption{Best-found shaping gain of convolutional code lattices formed by zero-tailed convolutional codes (ZTCCs) and tail-biting convolutional codes (TBCCs) for rate 1/2 and 1/3 with various memory orders $m$. The $0.65\dB$, $0.86\dB$ and $1.03\dB$ shaping gains of the $E_8$ lattice, the $BW_{16}$ lattice and the {Leech} lattice are also shown for comparison.}
\label{fig:cclsg}
\vspace{-10pt}
\end{figure}%
%\vspace{3cm}
\begin{table}[!t]
\centering
\renewcommand\arraystretch{1.1} %for double columns
    \caption{Recommended convolutional code generator polynomials for a range of dimensions $n$  based on best-found convolutional code lattices for shaping, and asymptotic shaping gain $\gamma_{\mathrm{s}}$ (octal number convention: $D^3+D+1$ is represented by 13).}
    \label{table:conv}
    \centering
  \resizebox{0.48\textwidth}{!}{% <------ Don't forget this % %for double columns
  %  \resizebox{0.6\textwidth}{!}{% <------ Don't forget this %
    \begin{tabular}{cccccccc}
    \toprule
\multicolumn{1}{c}{Convolutional code} &
\multicolumn{1}{c}{$m$} &
\multicolumn{1}{c}{$18\leq n\leq24$} &
\multicolumn{1}{c}{$24<n<72$} &
\multicolumn{1}{c}{$72\leq n \leq 144$} &
\multicolumn{1}{c}{$n > 144$} &
\multicolumn{1}{c}{asymptotic $\gamma_{\mathrm{s}}$ (dB)} &
\multicolumn{1}{c}{note} \\
 \toprule
\multirow{6}{5em}{Rate $1/2$, zero-tailed} &2 &$7,5$ &$7,5$ &$7,5$ &$7,5$ &0.9734 &- \\
  &3 &$17,13$ &$17,11$ &$17,13$ &$17,13$ &1.0622 &- \\
  &4 &$35,23$ &$33,25$ &$31,23$ &$31,23$ &1.1233 &$\Cconv_4$ \\
  &5 &$67,51$ &$77,55$ &$75,57$ &$75,57$ &1.1814 &$\Cconv_5$ \\
  &6 &$175,133$ &$175,133$ &$165,127$ &$165,127$ &1.2251 &- \\
  &7 &$365,327$ &$331,257$ &$357,251$ &$357,251$ &1.2574 &$\Cconv_6$ \\
\midrule
\multirow{4}{5em}{Rate $1/3$, zero-tailed}  &2 &$7,7,5$ &$7,7,5$ &$7,6,5$ &$7,6,5$ &0.9055 &$\Cconv_2$ \\
 &3 &$17,15,13$ &$17,15,13$ &$17,15,13$ &$17,15,13$ &1.0673 &- \\
 &4 &$37,33,25$ &$37,33,25$ &$37,33,25$ &$37,33,25$ &1.1321  &$\Cconv_3$ \\
 &5 &$71,65,57$ &$71,65,57$ &$73,57,41$ &$73,57,41$ &1.1808  &$\Cconv_1$ \\[0.4ex]
 \toprule
\multirow{6}{5em}{Rate $1/2$, tail-biting} &2 &$7,6$ &$7,5$ &$7,5$ &$7,5$ &0.9734 &- \\
  &3 &$16,3$ &$15,6$ &$17,13$ &$17,13$ &1.0622 &- \\
  &4 &$30,7$ &$30,13$ &$36,15$ &$31,23$ &1.1233 &-\\
  &5 &$70,3$ &$60,13$ &$74,13$ &$75,57$ &1.1814 &- \\
  &6 &$140,7$ &$140,13$ &$130,17$ &$165,127$ &1.2251 &-\\
  &7 &$340,3$ &$320,3$ &$320,17$ &$357,251$ &1.2574 &- \\
  \midrule
  \multirow{4}{5em}{Rate $1/3$,  tail-biting}  &2 &$7,6,4$ &$7,6,5$ &$7,6,5$ &$7,6,5$ &0.9055 &- \\
 &3 &$16,10,3$ &$13,10,7$ &$17,15,13$ &$17,15,13$ &1.0673 &- \\
 &4 &$30,10,7$ &$26,10,7$ &$36,26,23$ &$37,33,25$ &1.1321 &- \\
 &5 &$40,34,3$ &$70,13,10$  &$74,64,31$  &$73,57,41$ &1.1808 &- \\
% $\Cconv_1$ &$[73,57,41]$\\
% $\Cconv_2$  &[7,6,5]\\ 
% $\Cconv_3$  &[37,33,25]\\ 
% $\Cconv_4$  &[31,23]\\ 
% $\Cconv_5$  &[75,57]\\ 
% $\Cconv_6$ &[357,251]\\
 \bottomrule
    \end{tabular}%
}
\end{table} %the asymptotic shaping gain is obtained with at least 10^7 samples
%represented in octal numbers corresponding to the encoder implementation in a descending order
%\footnotetext{This convention is distinct from the representation in\cite[Ch.~12]{Lin-2004}.}
% \footnotetext{The octal generator polynomials in\cite[Sec.~V]{zhou-isita18} followed the same convention using descending order, but were incorrectly described when mentioned in\cite[Sec.~III-B]{zhou-isita18}. Also, this convention is distinct from the representation in\cite[Ch.~12]{Lin-2004}.}

Tail-biting convolutional codes have excellent coding performance at short-to-medium block length, thus are suitable to form {Construction A} shaping lattices for low-to-moderate dimension. The information length is $k=nR_{\textrm{TBCC}}$. 

A convolutional code lattice may be scaled by $K = 2^2,2^3,2^4,\ldots$ to be used with a {Construction D/D'} coding lattice to form a nested lattice code, so as to satisfy Lemma~\ref{lemma:sublattice}.

Generator polynomials which give good \textit{coding} properties for convolutional codes are well-known\cite[Ch.~12]{Lin-2004}. However, it is not clear if such generator polynomials are the best choice for \textit{shaping} lattices. We performed an exhaustive search of generator polynomials for rate $1/2$ and $1/3$ nonsystematic feedforward convolutional codes. For each one, the shaping gain of the resulting lattice was found by {Monte Carlo} integration using at least $10^7$ samples. 

For rate $1/2$ convolutional codes, it is worthwhile to mention that the generator polynomials for zero-tailed codes we found\footnote{We found these rate 1/2 code polynomials independently, and are grateful to the anonymous reviewer for pointing us to \cite{kudryashov-arxiv08,kudryashov-isit10}.  The shaping gains shown in\cite{kudryashov-arxiv08,kudryashov-isit10} are slightly higher, but by no more than $0.0066 \db$; we have no particular explanation for this discrepancy.} for asymptotic shaping gain match those provided in\cite{kudryashov-arxiv08}, except for $m=5$, where we found $(75,57)$ provides $0.01 \db$ higher asymptotic shaping gain than $(61,57)$. The shaping gain of tail-biting codes with short block length were also studied in\cite{kudryashov-isit10}, which is higher than that of the Leech lattice.

%Even though their shaping gains are higher, but with our simulations:
%our m=5 (75,57) provide 0.01 higher shaping gain than their (61,57)

The greatest shaping gain we found for various $m$ and $n$ is shown in~Fig.~\ref{fig:cclsg}. In general, tail-biting convolutional codes have higher shaping gains than zero-tailed convolutional codes, for a given dimension. For a range of dimensions, a generator polynomial with a shaping gain which is either the best-found shaping gain or within $0.01\dB$ to the best-found shaping gain is provided in Table~\ref{table:conv}, with exceptions as follows. An improvement for around $0.03\text{--}0.08\dB$ shaping gain can be obtained using generator polynomials $(77,76,73)$ at $n=18$ and $(331,257)$ at $n=24$ for zero-tailed convolutional codes, and using generator polynomials $(31,27)$, $(73,25)$, $(144,57)$, $(250,67)$, $(37,33,25)$ and $(75,45,26)$ at $n=144$ for tail-biting convolutional codes instead. The \emph{asymptotic shaping gain} obtained at $n=2^{20}$ and $n=2^{20}+2$ for rate $1/2$ and $1/3$ convolutional codes respectively is also provided. It is observed that at moderate dimensions the shaping gain of convolutional code lattices using tail-biting convolutional codes can achieve the asymptotic shaping gain.

\subsection{Complexity of Quantization}\label{subsection:complexity}

In this subsection, we study the tradeoff between shaping gain and quantization complexity for convolutional code lattices, when the Viterbi algorithm is used. {Construction A} lattice quantization\cite[p.~450]{Conway-1999} requires 5 operations per dimension to lift the binary codeword to a lattice point and the inverse. The {Viterbi} decoder uses $2^m$ comparisons at each trellis stage where the total number of trellis stages is $nR_{\textrm{ZTCC}}$. It is assumed that $n$ is much larger than $m$ so that the contribution of termination and initialization to complexity can be ignored. Thus the normalized time complexity is $5+2^mR_{\textrm{ZTCC}}$. 
%\footnote{The complexity increases exponentially as $m$ increases. The {Viterbi} algorithm is impractical when $m$ is large. When $m$ is small, the time complexity can be considered as nearly linear in $n$, that is $\mathcal{O}(n)$.}

%Lattices based on tail-biting convolutional codes have better shaping gain than that of using zero-tailed convolutional codes for short-to-moderate dimension, as shown in the previous subsection. 

%convolutional code lattices shaping gain versus time complexity
\begin{figure}[!t]
\centering
\includegraphics[width=0.48\textwidth]{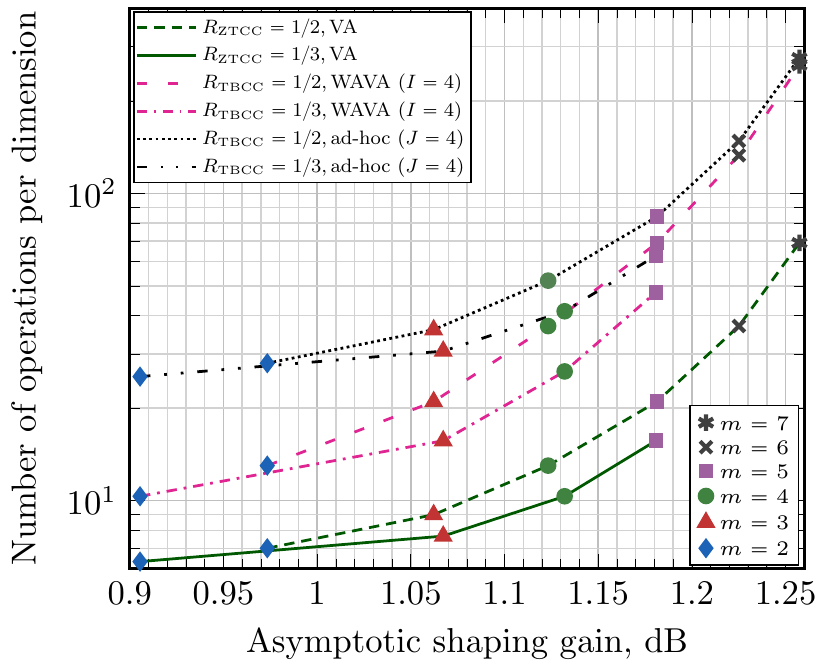} %for double columns
\caption{Performance-complexity tradeoff of convolutional code lattices formed by zero-tailed convolutional codes (ZTCCs) and tail-biting convolutional codes (TBCCs) with various memory orders $m$ where the decoding employs the {Viterbi} algorithm (VA) for ZTCCs, the wrap-around {Viterbi} algorithm (WAVA $I=4$ iterations) and the ad-hoc method ($J=4$ repeated times) for TBCCs.}
\label{fig:sgComplexity}
\end{figure}

We also analyzed the complexity of quantizing convolutional code lattices based on tail-biting convolutional codes using the wrap-around {Viterbi} algorithm\cite{Shao-com03} with a maximum of $I$ iterations and an ad-hoc suboptimal scheme\cite{Wang-pimrc96} that decodes repeated-$J$-times sequence using the {Viterbi} algorithm with zero termination, requiring $5+2^m R_{\textrm{TBCC}}I$ and $(5+2^m R_{\textrm{TBCC}})J$ operations per dimension respectively. The results given in the previous subsection were obtained using $J\geq 16$ and $nJ\geq1152$ for the ad-hoc decoding.

The normalized quantization complexity (or number of operations per dimension) is shown in~Fig.~\ref{fig:sgComplexity} as a function of asymptotic shaping gain. Rate $1/3$ convolutional codes outperform rate 1/2 convolutional codes for $m=3$ and $m=4$ in terms of shaping gain and quantization complexity, and convolutional code lattices based on rate 1/2 convolutional codes have the best shaping gain for a fixed memory order $m=2$ and $m=5$. Decoding tail-biting convolutional codes requires higher complexity than that of zero termination. In summary, using rate 1/3 convolutional codes produces a more favorable performance-complexity trade-off.  

The quantization for convolutional code lattices is optimal by employing the {Viterbi} algorithm and is close-to-optimal for tail-biting codes. The best-found efficiently achievable shaping gain is $1.25\dB$ with code rate $1/2$ and $m=7$ which is higher than the $1.03\dB$ of the {Leech} lattice.

\section{QC-LDPC {Construction D'} Lattices} \label{section:ldpcdesign}
In this section, we consider two-level {Construction D'} lattices. One approach of lattice construction can employ QC-LDPC codes and single parity-check product codes\cite{chen-istc18}. The first level code parity-check matrix consists of a top matrix that is modified from a QC-LDPC code\cite[Table~I]{rosnes-it12} and bottom rows which contribute to parity checks for the product code. The second level code parity-check matrix is constructed using row operations on a submatrix for the previous level's matrix. For this design, it is not clear how to obtain a triangular matrix for a {Construction D'} lattice. This work motivates us to design {Construction D'} lattices using only QC-LDPC codes where the second level code matrix $\H_1$ can be generated using row operations on a submatrix of the first level code matrix $\H_0$.

A design of QC-LDPC code $\C_0$ with a parity-check matrix $\H_0$ is presented, where the position of non-zero blocks is found by binary linear programming\cite{Zhou-isit21}. A \emph{subcode condition} $\C_0\subset\C_1$ must be satisfied to form a 2-level {Construction D'} lattice, and this is not straightforward. In\cite{da_silva-it19}, $\H_0$ was obtained from $\H_1$ by performing check splitting or PEG-based check splitting. In contrast to\cite{da_silva-it19} we design $\H_0$ so that $\H_1$ may be constructed using row operations, where $\H_0$ and $\H_1$ can be easily triangularized and thus efficient encoding and indexing is allowed. With this design, a straightforward method to find a triangular matrix for {Construction D'} lattices is also given.

\subsection{Design QC-LDPC Codes for {Construction D'}}\label{subsection:QCLDPCdesign}

The parity-check matrix $\H_0$ of a QC-LDPC code $\C_0$ can be expressed by
\begin{equation}
\H_0 =
\renewcommand\arraystretch{1.2} %for double columns
\begin{bmatrix}
\bP^{p^{}_{1,1}}		&\bP^{p^{}_{1,2}} 	   &\cdots 		 &\bP^{p^{}_{1,N}} 	\\
\bP^{p^{}_{2,1}}		&\bP^{p^{}_{2,2}} 	   &\cdots  	     &\bP^{p^{}_{2,N}} 	\\
\vphantom{\int}\smash[t]{\vdots} 		&\vphantom{\int}\smash[t]{\vdots} 	   &\vphantom{\int}\smash[t]{\ddots} 	     &\vphantom{\int}\smash[t]{\vdots} 	  \\
\bP^{p^{}_{M,1}}	&\bP^{p^{}_{M,2}}	   &\cdots 		 &\bP^{p^{}_{M,N}} 
\end{bmatrix},
\end{equation}
where $\bP$ is a $Z$-by-$Z$ right-shift cyclic-permutation matrix and $-1\leq p^{}_{i,j}<Z$ is an integer. For $i=1,2,\ldots,M$ and $j=1,2,\ldots,N$, when $p^{}_{i,j}=-1$, instead use the all-zeros matrix and $\mathbf P^0$ is the identity matrix $\I_Z$. The block length of $\C_0$ is $n=ZN$.

%----table----
\begin{table*}[!t]
%\setlength\belowcaptionskip{-0.5\baselineskip}
%\caption{Prototype matrix of $\H_0$ with circulant size $Z=96$ and block length $n=2304$. $*$ denotes a double circulant.}
\caption{Prototype matrix of $\H_0$ with $Z=96$ and $n=2304$ where $*$ denotes a double circulant}
\label{table:H0}
\centering
\renewcommand\arraystretch{1}
\resizebox{1\textwidth}{!}{% <------ Don't forget this %
\begin{tabular}{*{22}{r}cr}                                                                                                                                                    
-1 & -1 & 53 & -1 & 15 & 56 & -1 & -1 & 55 & 35 & -1 & 8 & 0 & -1 & -1 & -1 & -1 & -1 & -1 & -1 & -1 & -1 & -1 & -1 \\ 
                                                                                                                
-1 & -1 & 26 & -1 & -1 & 51 & -1 & 59 & 14 & -1 & 16 & -1 & 0 & 0 & -1 & -1 & -1 & -1 & -1 & -1 & -1 & -1 & -1 & -1 \\ 
                                                                                                                
18 & -1 & 3 & -1 & -1 & 82 & 42 & -1 & 33 & -1 & -1 & -1 & -1 & 0 & 0 & -1 & -1 & -1 & -1 & -1 & -1 & -1 & -1 & -1 \\  
                                                                                                                
-1 & 30 & 73 & 53 & -1 & 49 & -1 & -1 & 8 & -1 & -1 & -1 & -1 & -1 & 0 & 0 & -1 & -1 & -1 & -1 & -1 & -1 & -1 & -1 \\  
                                                                                                                
-1 & 67 & -1 & 15 & 84 & -1 & -1 & -1 & -1 & -1 & -1 & -1 & -1 & 3 & -1 & 82 & 0 & -1 & -1 & -1 & -1 & -1 & -1 & -1 \\ 
                                                                                                               
-1 & -1 & -1 & -1 & -1 & 71 & 83 & 34 & -1 & -1 & -1 & -1 & -1 & 0 & -1 & -1 & 25 & 0 & -1 & -1 & -1 & -1 & -1 & -1 \\ 
                                                                                                               
-1 & -1 & -1 & -1 & -1 & -1 & 8 & 27 & 87 & -1 & -1 & -1 & 0 & -1 & -1 & -1 & -1 & 59 & 0 & -1 & -1 & -1 & -1 & -1 \\  
                                                                                                               
-1 & -1 & -1 & -1 & -1 & -1 & -1 & -1 & 91 & -1 & 62 & 52 & -1 & -1 & -1 & 0 & -1 & -1 & 6 & 0 & -1 & -1 & -1 & -1 \\  
                                                                                                               
-1 & -1 & -1 & -1 & -1 & -1 & -1 & -1 & -1 & 11 & 5 & 17 & -1 & -1 & 0 & -1 & -1 & -1 & -1 & 12 & 0 & -1 & -1 & -1 \\  
                                                                                                               
-1 & -1 & 2 & 43 & 53 & -1 & -1 & -1 & -1 & -1 & -1 & -1 & -1 & -1 & 73 & -1 & -1 & -1 & -1 & -1 & 34 & 0 & -1 & -1 \\ 
                                                                                                                
54 & -1 & 26 & -1 & -1 & 12 & -1 & -1 & -1 & -1 & -1 & -1 & -1 & -1 & -1 & -1 & -1 & -1 & -1 & -1 & -1 & 9 &\ 0 & 0 \\                                                                                                                
52 & 91 & -1 & -1 & -1 & -1 & -1 & -1 & -1 & 38 & -1 & -1 & 13 & -1 & -1 & -1 & -1 & -1 & -1 & -1 & -1 & -1 & $66/71^*$  & 0 \\
\end{tabular}%
}
\end{table*}

%----table----
\begin{table*}[t]
%\setlength\belowcaptionskip{-0.5\baselineskip}
%\caption{Prototype matrix of $\H_1$ with circulant size $Z=96$ and block length $n=2304$. $*$ denotes a double circulant.}
\caption{Prototype matrix of $\H_1$ with $Z=96$ and $n=2304$ where $*$ denotes a double circulant}
\label{table:H1}
\centering
\renewcommand\arraystretch{1}
\resizebox{1\textwidth}{!}{% <------ Don't forget this %
\begin{tabular}{*{22}{r}cr}  
54 & 67 & 26 & 15 & 84 & 12 & 8 & 27 & 87 & 11 & 5 & 17 & 0 & 3 & 0 & 82 & 0 & 59 & 0 & 12 & 0 & 9 & 0 & 0 \\                                                                                                            
52 & 91 & 2 & 43 & 53 & 71 & 83 & 34 & 91 & 38 & 62 & 52 & 13 & 0 & 73 & 0 & 25 & 0 &6 & 0 & 34 & 0 & $66/71^*$ & 0 \\  
\end{tabular}%
}
\end{table*}                           
%----table----

Now we give a specific design of binary QC-LDPC codes $\C_0$ and $\C_1$ for 2-level {Construction D'} lattices. The parity-check matrices $\H_0$ and $\H_1$ are designed such that: 1) $\C_0 \subset \C_1$ 2) $\H_0$ and $\H_1$ are of full rank 3) $\H_0$ and $\H_1$ can be easily triangularized 4) $\H_0$ and $\H_1$ have girth as high as possible. Property 1) allows $\C_0$ and $\C_1$ to form a {Construction D'} lattice $\Lambda$. It is convenient to generate a triangular check matrix of $\Lambda$ using $\H_i$ with properties 2) and 3). Property 4) is designed subject to the error correction performance. 

To meet the design requirements, binary linear programming can be used to find a binary matrix with $M = 12$ rows and $N = 24$ columns whose element one represents a non-zero block in the prototype matrix of $\H_0$ and a zero represents a 0 block \cite{Zhou-isit21}. Then the prototype matrix of $\H_0$ is generated by choosing the $-1<p_{i,j}<Z$ of non-zero blocks such that the check matrices $\H_0$ and $\H_1$ have girth as high as possible, where $\H_1$ is constructed by the sum of the block rows of $\H_0$ in sets $\{5,7,9,11\}$ and $\{6,8,10,12\}$ respectively. For more detail, see\cite{Zhou-isit21}. For circulant size $Z=96$, the prototype matrix\footnote{The QC-LDPC code prototype matrices for $n=2304,5016,10008$ are available at https://github.com/fanzhou-code/qcldpc. The prototype matrix of $\H_0$ satisfies degree distribution modified from the structure in\cite[Table~I]{rosnes-it12} for variable nodes and check nodes: 
%Correction on Sep 13:
$\lambda(x) = \frac{7}{24}x^2+\frac{11}{24}x^3+\frac{1}{8}x^4+\frac{1}{8}x^6$ and $\rho(x)=\frac{7}{12}x^6+\frac{5}{12}x^7$, respectively, where $\lambda_dx^d$ and $\rho_dx^d$ means that $\lambda_d$ and $\rho_d$ are the node-perspective fraction of nodes with degree $d$. The prototype matrix of $\H_1$ has degree distribution polynomials $\lambda'(x) = \frac{23}{24}x^2+\frac{1}{24}x^3$ and $\rho'(x)=\frac{1}{2}x^{24}+\frac{1}{2}x^{25}$. The corresponding parity-check matrices $\H_0$ and $\H_1$ are of girth 8.} of $\H_0$ and $\H_1$ are given in Tables~\ref{table:H0}--\ref{table:H1}. Note that we assigned a double circulant $p^*_{12,23}=p^{(1)}_{12,23}, p^{(2)}_{12,23}$ such that $\H_0$ and $\H_1$ can be easily triangularized, which allows efficient encoding and indexing\cite{kurkoski-it18}. The design code rates of QC-LDPC codes are chosen similar to\cite{chen-istc18}, that is $1/2$ and $11/12$ for $\C_0$ and $\C_1$, respectively.

\subsection{Triangular Matrix of {Construction D'} Lattices}\label{subsection:trimatrixDprime}
A lower-triangular check matrix $\H$ for a 2-level {Construction D'} lattice is used for encoding. This can be constructed if the parity-check matrices $\H_0$ and $\H_1$ for nested binary codes $\C_0 \subset \C_1$ are triangularizable. Transform $\H_0$ and $\H_1$ into lower-triangular form by performing block row operations in the binary field, resulting in $\tH_0$ and $\tH_1$ respectively. The triangular matrix $\tH_0$ must contain the basis vectors of $\tH_1$ such that they both satisfy Definition~\ref{definition:nestedbincodes}. Then a lower-triangular check matrix\footnote{Although $\H$ obtained in this way introduces double circulants that might result in short cycles, this $\H$ is only used for encoding and indexing as described in Section~\ref{section:nestedlattice}. When decoding a Construction D' lattice as addressed in Subsection~\ref{subsection:decDprimelattice}, nontriangular matrices $\H_0$ and $\H_1$ are used by the binary decoders for $\C_0$ and $\C_1$, respectively.} $\H$ is built using Definition~\ref{definition:constructionDprime} in Section~\ref{section:Dprimelattice}. %Matrices used for decoding in Subsection~\ref{subsection:decDprimelattice} are $\H_0$ and $\H_1$, where re-encoding with respect to encoding method B in Subsection~\ref{subsection:encodeB} uses a lower-triangular unimodular square matrix $\tH=\begin{bsmallmatrix}\begin{smallmatrix}\I &\zero\end{smallmatrix}\\[-3.2pt]\tikz\draw[line width=0.05em,dashed] (0,0) -- (0.5,0);\\[-1.2pt] \tH_0\end{bsmallmatrix}$.}} $\H$ is built using Definition~\ref{definition:constructionDprime}. 

The design of parity-check matrices $\H_0$ and $\H_1$ for QC-LDPC codes given in the previous subsection allows a straightforward method to generate the lower-triangular check matrix $\tH_0$. Let $p^{(1)}_{12,23} = \fa$ and $p^{(2)}_{12,23} = \fb$ be selected such that
\begin{align}
\bQ =\I_Z + \bP^{\fa} + \bP^{\fb}		\label{eqn:matrixQ}
\end{align}
is a triple circulant and full rank. The lower-triangular $\tH_1$ can be obtained as follows. Let $\bV$ be the block-wise sum of the two block rows of $\H_1$ over GF(2). The twenty-third block column of $\bV$ is a square matrix $\bQ$~(\ref{eqn:matrixQ}). Using only row operations over GF(2), $\bQ$ can be transformed to triangular form $\T$. Find a binary matrix $\bW$ such that $\bW \odot \bQ = \T$. Replace the first block row of $\H_1$ by $\bW \odot \bV$ then the resulting matrix is lower-triangular and denoted $\tH_1$. After that, $\tH_0$ is built by replacing the bottom two block rows of $\H_0$ by $\tH_1$.

\section{Numerical Results} \label{section:simulationresult}

{Construction D'} lattices $\Lc$ of dimension $n=2304,5016,10008$ formed by QC-LDPC codes were evaluated in the power-constrained AWGN channel. At the decoder, the re-encoding implicitly assumes that method B of Section~\ref{section:Dprimelattice} is being used, which is equivalent to method A of Section~\ref{section:Dprimelattice}, even for triangular Construction D' matrices of Section~\ref{section:ldpcdesign}. The belief propagation decoder of LDPC codes ran maximum $50$ iterations. The well-known low-dimensional $E_8$, $BW_{16}$ and {Leech} lattices were each used for shaping a $2304$-dimensional coding lattice. We also used convolutional code lattices for shaping. A variety of zero-tailed convolutional codes were chosen based on the best-found generator polynomials and complexity analysis of quantization employing the Viterbi algorithm in Section~\ref{section:convshaping}, for shaping $n=2304,5016,10008$-dimensional {Construction D'} lattices. The channel model follows~Fig.~\ref{fig:blockdiagramnestedlattice} where the encoding and indexing are performed as shown in Subsection~\ref{subsection:encodeindex}.%The numerical results are given as follows.

For comparison we performed hypercube shaping\footnote{The work in\cite{da_silva-it19} can also produce a hypercubical constellation, but we perform hypercube shaping with respect to our proposed decoding algorithm.} where lattice points of an $a$-level Construction D' lattice were transformed into a hypercube $\B = \{0,1,\ldots,L-1\}^n$ for an integer $L$ being a multiple of $2^a$. Hypercube shaping for Construction D' can be performed as follows. Let a Construction D' lattice $\Lc$ have a lower triangular check matrix $\Hc$ with diagonal elements $h_{i,i}$ for $i=1,\ldots,n$, and let $L\I_n$ be a generator matrix of the ``shaping lattice'' $\Ls=L\Zn$ where $\I_n$ is an identity matrix of size $n$. Choose $L$ such that the product of $\Hc$ and $L\I_n$ is a matrix of integers. The information vector consists of integers in $\{0,1,\ldots,Lh_{i,i}-1\}$. Performing modulo-$L$ on a lattice point of $\Lc$ is the ``shaping'' operation reducing the lattice point in a hypercube $\B$. This is simpler than the sequential computations in\cite[eq.~(1)-(3)]{Sommer-itw09} to transform a lattice point into $\B$.
%This can be done for a {Construction D'} lattice point by performing modulo-$L$, instead of computing sequentially in\cite[eq.~(1)-(3)]{Sommer-itw09}. 
The code rate is $R' = \frac{1}{n}\log_2\frac{L^n}{|\det(\Gc)|}$. 

For $a$-level Construction D' lattices with hypercube shaping, it is natural to use $2^a$-PAM signalling.  For the shaped lattice codes in this paper, the lattice points $\x$ are integers due the use of Construction D'; however greater than $2^a$ modulation levels are required. Construction D' lattices with hypercube shaping can also use greater than $2^a$ modulation levels, but no shaping gain is provided.

%high rate shaping Construction D' lattices, and hypercube shaping (Eb/N0)
%$E_8$, $BW_16$, Leech and convolutioanl code lattice shaping
\begin{figure}[!t]
\centering
\includegraphics[width=0.45\textwidth]{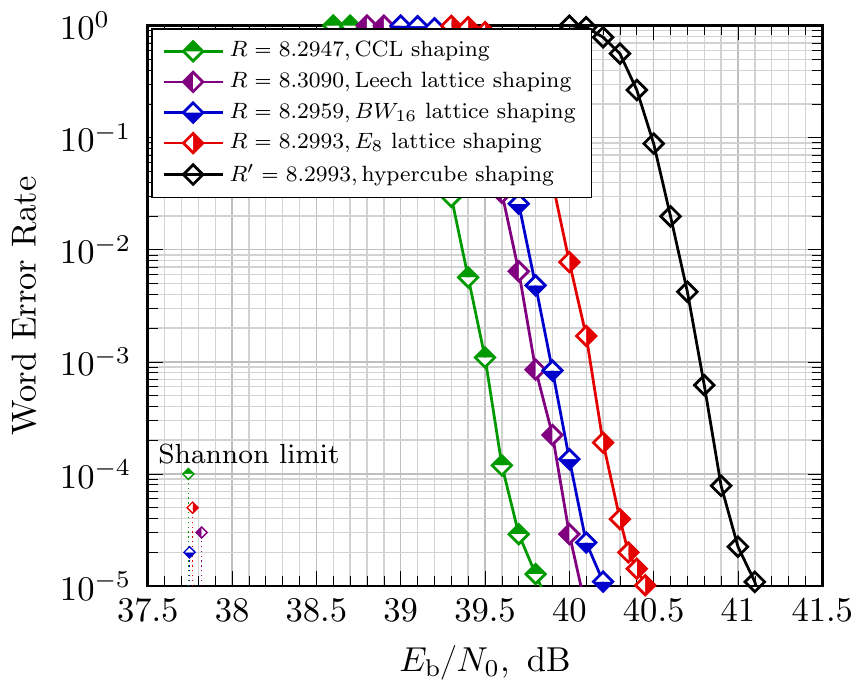} %for double columns
\caption{Word error rate as a function of $\ebnoshort$ using a variety of lattices for shaping a 2304-dimensional {Construction D'} lattice, where the convolutional code lattice (CCL) is formed by a zero-tailed convolutional code $\Cconv_6$ with $128$ states.}% Hypercube shaping was performed for comparison. }
\label{fig:shapingresulthighrate}
\end{figure}
 
%convolutional code lattice shaping Construction D' lattices, and hypercube shaping (Eb/N0)
\begin{figure}[!pt]
\centering
\includegraphics[width=0.45\textwidth]{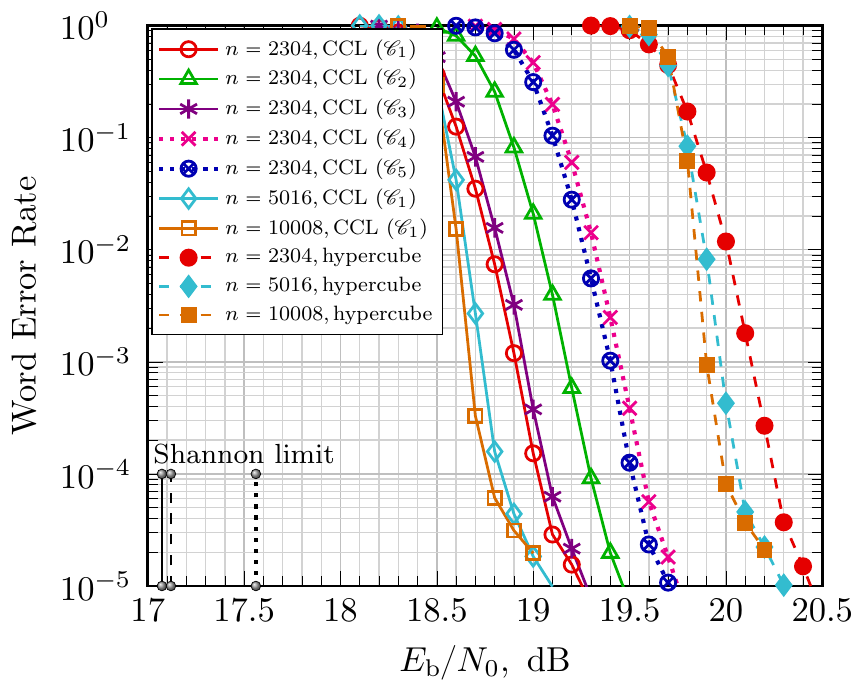} %for double columns
\caption{Word error rate as a function of $\ebnoshort$ using various convolutional code lattices (CCLs) based on $\Cconv_1$-$\Cconv_5$ with generator polynomials in~Table~\ref{table:conv} for shaping $n$-dimensional {Construction D'} lattices where the code rate is listed in~Table~\ref{table:coderate}.} %The code rate of using CCL shaping is slightly smaller than that of hypercube shaping. 
%See Table~\ref{table:coderate}.}
\label{fig:convoshapingresult}
\end{figure}%

\subsection{\texorpdfstring{$E_8$, $BW_{16}$}{E8,BW16} and {Leech} Lattice Shaping}
 \label{subsection:lowdimlatticeshaping}
Well-known low-dimensional lattices were used for shaping high-dimensional lattices because they can provide good shaping gains and their decoding is well-studied. The $E_8$ lattice, the $BW_{16}$ lattice and the {Leech} lattice have optimal quantization algorithms\cite{Conway-it82,Conway-siam84,Viterbo-it99}. The authors in\cite{Ferdinand-twc16} used the $E_8$ and $BW_{16}$ lattices for shaping LDLC lattices. At $n=24$ the {Leech} lattice has a shaping gain of $1.03\dB$, which was used for shaping LDA lattices\cite{di_pietro-com17}. Following\cite{Ferdinand-twc16,di_pietro-com17} we built shaping lattices using direct sum of scaled copies of the $E_8$, $BW_{16}$, and {Leech} lattices by a scale factor $K$. Let $\Hc$ be the check matrix of an $n$-dimensional {Construction D'} coding lattice, and $\G$ be the generator matrix of an $n'$-dimensional lattice where $n/n'$ is an integer. The factor $K$ is chosen such that $\Hc\Gs \in \Zn$ where $\Gs$ is a block diagonal matrix of size $n/n'$ with each block $K\G$. %The shaping lattice quantizer uses decoder of the $E_8$ and $BW_{16}$ lattices, or a sphere decoder for {Leech} lattice. 
Rectangular encoding and its inverse indexing can be efficiently implemented due to the lower-triangular structure in matrix $\Hc$ and $\Gs$. By choosing various $K$ we generated nested lattice codes with a variety of code rates $R$. 
%A maximum shaping gain of $0.65\db$, $0.86\dB$ and $1.03\dB$ can be provided using $E_8$, $BW_{16}$ and {Leech} lattice shaping.

For shaping the $2304$-dimensional {Construction D'} lattice, the same code rate for both the $E_8$ lattice shaping and hypercube shaping can be easily achieved. Let $K_{BW_{16}} = 280\sqrt{2}$ and $K_{\textrm{Leech}} = 168\sqrt{8}$, then $BW_{16}$ and {Leech} lattice shaping produce code rate approximately $8.2959$ and $8.3090$, respectively, close to $R=R'=8.2993$ of choosing $K_{E_8} = L = 472$. The word error rate is given in~Fig.~\ref{fig:shapingresulthighrate} as a function of $\ebnoshort$. If we take account of the code rate differences, a $0.65\dB$, $0.86\dB$ and $1.03\dB$ shaping gain is preserved respectively, as the full shaping gain of the $E_8$, $BW_{16}$ and {Leech} lattices.

\subsection{Convolutional Code Lattices for Shaping {Construction D'} Lattices} \label{subsection:convshapingresult}
In this paper we consider high-dimensional {Construction D'} lattices, thus zero-tailed convolutional codes are suitable for constructing convolutional code lattices for shaping. At $n\geq2304$, using zero-tailed convolutional codes provides comparable shaping gain and requires lower quantization complexity than that of tail-biting convolutional codes. A variety of convolutional code lattices based on rate $1/2, 1/3$ zero-tailed convolutional codes selected from Table~\ref{table:conv} were also used for shaping the proposed QC-LDPC {Construction D'} Lattices, where the smallest possible scale factor $K=4$ to satisfy Lemma~\ref{lemma:sublattice} can produce a code rate approximately $2.084$ and $1.917$ respectively.

Lattices are ideal at high code rate thus we chose $K>4$ for evaluation. The nested lattice code parameters in our simulations are listed in Table~\ref{table:coderate}, including the code rates, close to that of hypercube shaping for a fair comparison. The numerical results in terms of word error rate as a function of $\ebnoshort$ are shown in~Fig.~\ref{fig:convoshapingresult}. Convolutional code lattice shaping using a rate 1/3 convolutional code with $m=5$ was performed for $n=2304, 5016, 10008$, showing an improvement on the error-correction performance and the shaping gain as $n$ increases. For a fixed dimension $n=2304$, we show that a higher shaping gain is achieved by increasing the memory order $m$. The numerical results of using rate 1/2 zero-tailed convolutional codes are also provided, where the code rate was chosen as close as possible to hypercube shaping. The resulting shaping gains are approximate to the estimated shaping gains listed in Table~\ref{table:coderate} if we take account of the code rate differences.

\begin{table}[!pt]
\renewcommand\arraystretch{1.1} %for double columns
    \caption{Code rate $R$ of nested lattice codes using various convolutional code $\Cconv$ with memory order $m$, where the convolutional code lattice is scaled by a factor $K$. The Monte Carlo estimated shaping gain $\gamma_{\mathrm{s}}$ is given in decibels. Hypercube side length $L$ is chosen to achieve $R'\approx R$}
    \label{table:coderate}
    \centering
    %\resizebox{0.45\textwidth}{!}{% <------ Don't forget this % %for double columns
   \resizebox{0.45\textwidth}{!}{% <------ Don't forget this %
    \begin{tabular}{ccclclcc}
    \toprule
\multicolumn{1}{c}{Dimension} &
\multicolumn{5}{c}{Convolutional code lattice shaping}    &
\multicolumn{2}{c}{Hypercube shaping}    \\ 
\cmidrule(lr){1-1}
\cmidrule(lr){2-6}
\cmidrule(lr){7-8}
\multicolumn{1}{c}{$n$} &
\multicolumn{1}{c}{\quad$m$} &
\multicolumn{1}{c}{$\Cconv$} &
\multicolumn{1}{c}{$\gamma_{\mathrm{s}}$ (dB)} &
\multicolumn{1}{c}{$\quad K$} &
\multicolumn{1}{c}{$R$} &
\multicolumn{1}{c}{\quad $L$} &
\multicolumn{1}{c}{$R'$} \\
\toprule
         2304    &\quad 5  &$\Cconv_1$ &1.1731 &\quad 20 &4.4074     &\quad 32 &4.4167 \\
         5016    &\quad 5  &$\Cconv_1$ &1.1772 &\quad 20 &4.4063     &\quad 32 &4.4167 \\
         10008   &\quad 5  &$\Cconv_1$ &1.1790 &\quad 20 &4.4058     &\quad 32 &4.4167 \\ \midrule
         2304    &\quad 2  &$\Cconv_2$ &0.9022 &\quad 20 &4.4061     &\quad 32 &4.4167 \\
         2304    &\quad 4  &$\Cconv_3$ &1.1259 &\quad 20 &4.4070     &\quad 32 &4.4167 \\ 
         2304    &\quad 4  &$\Cconv_4$ &1.1186 &\quad 24 &4.5034     &\quad 32    &4.4167 \\
         2304    &\quad 5  &$\Cconv_5$ &1.1756 &\quad 24  &4.5038     &\quad 32    &4.4167 \\
         \midrule
         2304       &\quad 7  &$\Cconv_6$ &1.2500 &\quad 332 &8.2947    &\quad 472    &8.2993 \\
         \bottomrule
    \end{tabular}%
}
\end{table}
%more accurate shaping gain is used for at least 10^7 samples

We also compared the shaping gain of a convolutional code lattice (chose $\Cconv_6$ to produce a high shaping gain) with that of the $E_8$, $BW_{16}$ and {Leech} lattices as plotted in~Fig.~\ref{fig:shapingresulthighrate} for $n=2304$. The shaping gain of $1.25\dB$ was preserved with convolutional code lattice shaping---this is the best-found shaping gain achieved by lattice shaping in the power-constrained channel, to the best of the authors' knowledge. For the four shaping lattices: convolutional code lattice, the $E_8$ lattice, the $BW_{16}$ lattice and the {Leech} lattice, using a smallest possible scale factor $4,4,4\sqrt{2},4\sqrt{8}$ respectively for shaping the proposed 2304-dimensional {Construction D'} lattice, the integers solutions $e_i\in[0,\beta)$ (\ref{eqn:indexsolvee}) are bounded by $\beta=8,16,16,32$. The values of integers are bounded by $\beta=944,944,1120,1344$ for the results in Fig.~\ref{fig:shapingresulthighrate}. Regarding the distance to the Shannon limit, while the LDA lattice construction\cite{di_pietro-com17} has better performance, it requires nonbinary LDPC codes, whereas our construction uses lower-complexity binary LDPC codes. The LDLC construction\cite{Ferdinand-twc16} has similar performance, but higher decoding complexity than binary LDPC codes.

\section{Conclusion}\label{section:conclusion}
This paper addressed the problem of encoding and decoding of {Construction D'} lattices for the power-constrained channel. Our lattice constructions provide both good coding properties and efficiently achievable high shaping gains. From a practical point of view, they are also suitable for hardware implementations, as well-understood QC-LDPC codes and convolutional codes with the Viterbi algorithm are used. An open problem is to optimize the LDPC degree distributions using density evolution techniques, which help us design QC-LDPC codes with good error-correction performance, such that the resulting QC-LDPC {Construction D'} lattices are optimized.

\ifCLASSOPTIONcaptionsoff
  \newpage
\fi



%\setstretch{1.15}
% %\bibliographystyle{bibtex/IEEEtranTCOM} 
\bibliographystyle{bibtex/IEEEtrannourl}
% % argument is your BibTeX string definitions and bibliography database(s)
% \bibliography{bibtex/IEEEabrv,bibtex/abbrev,bibtex/bits-conf,bibtex/bits-invited,bibtex/bits-journal,bibtex/bits-noreview,
% bibtex/bits-recent,bibtex/bits,bibtex/bitsbib,bibtex/starbib,bibtex/thisbib}


%\begin{biography}[{\includegraphics[width=1in,height=1.25in,clip,keepaspectratio]{mshell}}]{Michael Shell}


%\vfill


\end{document}

%% file: preamble/myLatexCommands.tex
\usepackage{graphicx,amssymb,amsfonts,adjustbox,arydshln,mathtools,balance,cite}
\usepackage{eqnarray}
\usepackage{mathabx,algorithmic,setspace}
\usepackage[ruled]{algorithm2e}
\usepackage{booktabs, makecell, multirow, tabularx}
\usepackage{mathrsfs}

\usepackage{tikz,pgf,pgffor,pgfplots,pgfplotstable,subfigure}
\usetikzlibrary{shapes,arrows.meta,positioning,shapes.geometric,fit,backgrounds,matrix,decorations.pathreplacing,calc}
\pgfplotsset{compat=1.17} 

\pgfkeys{tikz/mymatrixenv/.style={decoration=brace,every left delimiter/.style={xshift=3pt},every right delimiter/.style={xshift=-3pt}}}
\pgfkeys{tikz/mymatrix/.style={matrix of math nodes,left delimiter=[,right delimiter={]},inner sep=2pt,column sep=1em,row sep=0.5em,nodes={inner sep=0pt}}}
\pgfkeys{tikz/mymatrixbrace/.style={decorate,thick}}
\newcommand\mymatrixbraceoffseth{0.5em}

\newcommand*\mymatrixbraceleft[4][m]{
    \draw[mymatrixbrace] ($(#1.north east)!(#1-#2-1.north east)!(#1.south east)+(\mymatrixbraceoffseth,0)$)
        -- node[right=2pt] {#4} 
        ($(#1.north east)!(#1-#3-1.south east)!(#1.south east)+(\mymatrixbraceoffseth,0)$);
}

\tikzset{
    cheating dash/.code args={on #1 off #2}{
        \csname tikz@addoption\endcsname{%
            \pgfgetpath\currentpath%
            \pgfprocessround{\currentpath}{\currentpath}%
            \csname pgf@decorate@parsesoftpath\endcsname{\currentpath}{\currentpath}%
            \pgfmathparse{\csname pgf@decorate@totalpathlength\endcsname-#1}\let\rest=\pgfmathresult%
            \pgfmathparse{#1+#2}\let\onoff=\pgfmathresult%
            \pgfmathparse{max(floor(\rest/\onoff), 1)}\let\nfullonoff=\pgfmathresult%
            \pgfmathparse{max((\rest-\onoff*\nfullonoff)/\nfullonoff+#2, #2)}\let\offexpand=\pgfmathresult%
            \pgfsetdash{{#1}{\offexpand}}{0pt}}%
    }
}

\usepackage{xpatch}
\makeatletter
\xpatchcmd{\@thm}{\thm@headpunct{.}}{\thm@headpunct{}}{}{}
\makeatother

\makeatletter
\def\bstctlcite{\@ifnextchar[{\@bstctlcite}{\@bstctlcite[@auxout]}}
\def\@bstctlcite[#1]#2{\@bsphack
  \@for\@citeb:=#2\do{%
    \edef\@citeb{\expandafter\@firstofone\@citeb}%
    \if@filesw\immediate\write\csname #1\endcsname{\string\citation{\@citeb}}\fi}%
  \@esphack}
\makeatother
\mathchardef\mhyphen="2D

\DeclareMathOperator{\lcm}{lcm\thinspace}

\newcommand{\C}{\mathcal C}
\newcommand{\V}{\mathcal V}

\newcommand{\B}{\mathcal B}

\newcommand{\Rn}{\mathbb{R}^n}
\newcommand{\F}{\mathbb F}

\newcommand{\Zn}{\mathbb{Z}^n}

\newcommand{\fC}{\mathfrak C}
\newcommand{\Cconv}{\mathscr C}

\renewcommand{\H}{\mathbf H}

\newcommand{\G}{\mathbf G}
\newcommand{\D}{\mathbf D}
\newcommand{\T}{\mathbf T}

\newcommand{\fa}{\mathfrak a}
\newcommand{\fb}{\mathfrak b}

\newcommand{\mA}{\mathbf A}
\newcommand{\mB}{\mathbf B}
\newcommand{\mC}{\mathbf C}

\newcommand{\mX}{\mathbf X}

\newcommand{\LA}{\Lambda_{\mathrm A}}
\newcommand{\GLA}{\G_{\LA}}

\newcommand{\bb}{\mathbf b}

\newcommand{\e}{\mathbf e}
\newcommand{\bu}{\mathbf u}

\newcommand{\g}{\mathbf g}
\newcommand{\bt}{\mathbf t}
\newcommand{\h}{\mathbf h}
\newcommand{\w}{\mathbf w}
\newcommand{\s}{\mathbf s}
\newcommand{\x}{\mathbf x}
\newcommand{\y}{\mathbf y}
\newcommand{\z}{\mathbf z}
\newcommand{\I}{\mathbf I}

\newcommand{\zero}{\mathbf 0}
\newcommand{\bP}{\mathbf P}
\newcommand{\bQ}{\mathbf Q}
\newcommand{\bU}{\mathbf U}
\newcommand{\bV}{\mathbf V}
\newcommand{\bW}{\mathbf W}

\newcommand{\Hc}{\mathbf{H}_{\mathrm{c}}}

\newcommand{\Gc}{\mathbf{G}_{\mathrm{c}}}
\newcommand{\Gs}{\mathbf{G}_{\mathrm{s}}}
\newcommand{\Ls}{\Lambda_{\mathrm{s}}}
\newcommand{\Lc}{\Lambda_{\mathrm{c}}}

\newcommand{\thx}{\widehat{\widetilde{\mathbf x}}}
\newcommand{\tx}{\widetilde{\mathbf x}}

\newcommand{\hx}{\widehat{\mathbf x}}
\newcommand{\tH}{\widetilde{\mathbf H}}
\newcommand{\hu}{\widehat{\mathbf u}}

\newcommand{\starmod}{\textrm{mod}^*\thinspace}

\newcommand{\modtwo}{\textrm{mod}_2\thinspace}
\renewcommand{\mod}{\textrm{mod}\thinspace}

\newcommand{\Dec}{\textrm{Dec}}

\renewcommand{\det}{\textrm{det}\thinspace}

\newcommand{\SNR}{\textrm{SNR}}

\newcommand{\dB}{\textrm{ dB}}
\newcommand{\db}{\textrm{ dB}}

\newcommand{\es}{E_{\mathrm s}}
\newcommand{\eb}{E_{\mathrm b}}
\newcommand{\no}{N_0}

\newcommand{\ebnoshort}{\eb/\no}
\newcommand{\var}{\sigma^2}

\newtheorem{definition}{Definition}
\newtheorem{lemma}{Lemma}
\newtheorem{proposition}{Proposition}

\newtheorem{example}{Example}
\renewcommand{\IEEEQED}{\IEEEQEDopen}

%% file: Zhou-TCOM.bbl
\begin{thebibliography}{10}
\providecommand{\url}[1]{#1}
\csname url@samestyle\endcsname
\providecommand{\newblock}{\relax}
\providecommand{\bibinfo}[2]{#2}
\providecommand{\BIBentrySTDinterwordspacing}{\spaceskip=0pt\relax}
\providecommand{\BIBentryALTinterwordstretchfactor}{4}
\providecommand{\BIBentryALTinterwordspacing}{\spaceskip=\fontdimen2\font plus
\BIBentryALTinterwordstretchfactor\fontdimen3\font minus
  \fontdimen4\font\relax}
\providecommand{\BIBforeignlanguage}[2]{{%
\expandafter\ifx\csname l@#1\endcsname\relax
\typeout{** WARNING: IEEEtran.bst: No hyphenation pattern has been}%
\typeout{** loaded for the language `#1'. Using the pattern for}%
\typeout{** the default language instead.}%
\else
\language=\csname l@#1\endcsname
\fi
#2}}
\providecommand{\BIBdecl}{\relax}
\BIBdecl

\bibitem{Forney-it98}
G.~D.~Forney and G.~Ungerboeck, ``Modulation and coding for linear
  {Gaussian} channels,'' \emph{IEEE Transactions on Information Theory},
  vol.~44, no.~6, pp.~2384--2415, Oct.~1998.
\BIBentrySTDinterwordspacing

\bibitem{bocherer-com15}
G.~B{\"{o}}cherer, F.~Steiner, and P.~Schulte, ``Bandwidth
  efficient and rate-matched low-density parity-check coded modulation,'' \emph{IEEE Transactions on Communications}, vol.~63,
  no.~12, pp.~4651--4665, Dec.~2015.
\BIBentrySTDinterwordspacing

\bibitem{Nazer-it11}
B.~Nazer and M.~Gastpar, ``Compute-and-forward: {Harnessing} interference through structured codes,'' \emph{IEEE Transactions on Information Theory}, vol.~57, no.~10, pp.~6463--6486, Oct.~2011.
\BIBentrySTDinterwordspacing

\bibitem{Conway-it83}
J.~H.~Conway and N.~J.~A.~Sloane, ``A fast encoding method for lattice codes and quantizers,'' \emph{IEEE Transactions on Information Theory}, vol.~29, no.~6, pp.~820--824, Nov.~1983.
\BIBentrySTDinterwordspacing

\bibitem{Forney-jsac89*2}
G.~D.~Forney, ``Multidimensional constellations---{Part II}: {Voronoi}
  constellations,'' \emph{IEEE Journal on Selected Areas in Communications}, vol.~7, no.~6, pp.~941--958, Aug.~1989.
\BIBentrySTDinterwordspacing

\bibitem{Erez-it04}
U.~Erez and R.~Zamir, ``Achieving $\frac{1}{2}\log\, (1+ \textrm{SNR})$ on the {AWGN} channel with lattice encoding and decoding,'' \emph{IEEE Transactions on Information Theory}, vol.~50, no.~10, pp.~2293--2314, Oct.~2004.
\BIBentrySTDinterwordspacing

\bibitem{di_pietro-it18}
N.~di~Pietro, G.~Z{\'e}mor, and J.~J.~Boutros, ``{LDA}
  lattices without dithering achieve capacity on the {Gaussian} channel,'' \emph{IEEE Transactions on Information Theory}, vol.~64, no.~3, pp.~1561--1594, Mar.~2018.
\BIBentrySTDinterwordspacing

\bibitem{kurkoski-it18}
B.~M.~Kurkoski, ``Encoding and indexing of lattice codes,'' \emph{IEEE
  Transactions on Information Theory}, vol.~64, no.~9, pp.~6320--6332,
  Sep.~2018.
\BIBentrySTDinterwordspacing

\bibitem{Sadeghi-it06}
M.-R.~Sadeghi, A.~H.~Banihashemi, and D.~Panario, ``Low-density parity-check
  lattices: construction and decoding analysis,'' \emph{IEEE Transactions on Information Theory}, vol.~52, no.~10, pp.~4481--4495, Oct.~2006.
\BIBentrySTDinterwordspacing

\bibitem{da_silva-it19}
P.~R.~Branco~da~Silva and D.~Silva, ``Multilevel {LDPC}
  lattices with efficient encoding and decoding and a generalization of
  {Construction~D'},'' \emph{IEEE Transactions on Information
  Theory}, vol.~65, no.~5, pp.~3246--3260, May~2019.
\BIBentrySTDinterwordspacing

\bibitem{Conway-1999}
J.~H.~Conway and N.~J.~A.~Sloane, \emph{Sphere Packings, Lattices and Groups}, 3rd~ed.\hskip 1em plus 0.5em minus 0.4em\relax New York, NY, USA:
  Springer-Verlag, 1999.
\BIBentrySTDinterwordspacing

\bibitem{Zamir-2014}
R.~Zamir, \emph{Lattice Coding for Signals and Networks}.\hskip 1em plus 0.5em minus 0.4em\relax Cambridge, UK: Cambridge, 2014.
\BIBentrySTDinterwordspacing

\bibitem{chen-istc18}
S.~Chen, B.~M.~Kurkoski, and E.~Rosnes,
  ``{Construction~D'} lattices from quasi-cyclic
  low-density parity-check codes,'' in \emph{Proc.~IEEE 10th International Symposium on Turbo
  Codes~\&~Iterative Information Processing}, 2018, pp.~1--5.
\BIBentrySTDinterwordspacing



\bibitem{Erez-it05*3}
U.~Erez and S.~ten~Brink, ``A close-to-capacity dirty paper coding scheme,'' \emph{IEEE Transactions on Information Theory}, vol.~51, no.~10, pp.~3417--3432, Oct.~2005.
\BIBentrySTDinterwordspacing

\bibitem{kudryashov-arxiv08}
B.~Kudryashov and K.~Yurkov, ``Linear code-based vector
  quantization for independent random variables,'' 2008, arXiv:0805.2379
  [cs.IT].
\BIBentrySTDinterwordspacing

\bibitem{kudryashov-isit10}
B.~D. Kudryashov and K.~V. Yurkov, ``Near-optimum
  low-complexity lattice quantization,'' in
  \emph{Proc.~2010 IEEE International Symposium on Information Theory}, 2010, pp.~1032--1036.
\BIBentrySTDinterwordspacing



\bibitem{Sommer-itw09}
N.~Sommer, M.~Feder, and O.~Shalvi, ``Shaping methods for low-denisty lattice codes,'' in \emph{Proc.~2009 IEEE Information Theory Workshop}, 2009, pp.~238--242.
\BIBentrySTDinterwordspacing

\bibitem{khodaiemehr-com17}
H.~Khodaiemehr, M.-R.~Sadeghi, and A.~Sakzad, ``Practical encoder and decoder for power constrained {QC} {LDPC}-lattice codes,'' \emph{IEEE Transactions on Communications}, vol.~65, no.~2, pp.~486--500, Feb.~2017.
\BIBentrySTDinterwordspacing

\bibitem{Ferdinand-twc16}
N.~S.~Ferdinand, B.~M.~Kurkoski, M.~Nokleby, and B.~Aazhang, ``Low-dimensional shaping for high-dimensional lattice codes,'' \emph{IEEE Transactions on Wireless Communications}, vol.~15, no.~11, pp.~7405--7418, Nov.~2016.
\BIBentrySTDinterwordspacing

\bibitem{di_pietro-com17}
N.~di~Pietro and J.~J.~Boutros, ``{Leech} constellations of {Construction-A} lattices,'' \emph{IEEE Transactions on Communications}, vol.~65, no.~11, pp.~4622--4631, Nov.~2017.
\BIBentrySTDinterwordspacing


\bibitem{Zhou-commlett17}
F.~Zhou and B.~M. Kurkoski, ``Shaping {LDLC} lattices using convolutional code lattices,'' \emph{IEEE Communications Letters}, vol.~21, no.~4, pp.~730--733, Apr.~2017.
\BIBentrySTDinterwordspacing

\bibitem{buglia-commlett21}
H.~Buglia and R.~R.~Lopes, ``{Voronoi} shaping for lattices with efficient encoding,'' \emph{IEEE Communications Letters}, vol.~25, no.~5, pp.~1439--1442, May~2021.
\BIBentrySTDinterwordspacing



\bibitem{richardson-it01*3}
T.~J.~Richardson and R.~L.~Urbanke, ``Efficient encoding of
  low-density parity-check codes,'' \emph{IEEE
  Transactions on Information Theory}, vol.~47, no.~2, pp.~638--656, Feb.~2001.
\BIBentrySTDinterwordspacing

\bibitem{ouellette-laa81}
D.~V.~Ouellette, ``Schur complements and statistics,'' \emph{Linear Algebra and its Applications}, vol.~36, pp.~187--295, Mar.~1981.
\BIBentrySTDinterwordspacing

\bibitem{vem-isit14}
A.~Vem, Y.-C.~Huang, K.~R.~Narayanan, and H.~D.~Pfister,
  ``Multilevel lattices based on spatially-coupled {LDPC} codes with applications,'' in \emph{Proc.~2014 IEEE International Symposium on Information Theory}, 2014, pp.~2336--2340.
\BIBentrySTDinterwordspacing

% \bibitem{Sakzad-aller10}
% A.~Sakzad, M.~Sadeghi, and D.~Panario, ``Construction of turbo lattices,'' in \emph{Proc.~48th Annual Allerton Conference on Communication, Control, and Computing}, 2010, pp.~14--21.
% \BIBentrySTDinterwordspacing

% \bibitem{Yan-itw12}
% Y.~Yan and C.~Ling, ``A construction of lattices from polar codes,'' in \emph{Proc.~2012 IEEE Information Theory Workshop}, 2012, pp.~124--128.
% \BIBentrySTDinterwordspacing

% \bibitem{matsumine-glocom18}
% T.~Matsumine, B.~M.~Kurkoski, and H.~Ochiai,
%   ``{Construction~D} lattice decoding and its
%   application to {BCH} code lattices,'' in
%   \emph{Proc.~2018 IEEE Global Communications Conference}, 2018, pp.~1--6.
% \BIBentrySTDinterwordspacing

\bibitem{Conway-siam84}
J.~H.~Conway and N.~J.~A.~Sloane, ``On the {Voronoi} regions of certain lattices,'' \emph{{SIAM} Journal on Algebraic Discrete Methods}, vol.~5, no.~3, pp.~294--305, Sep.~1984.
\BIBentrySTDinterwordspacing

\bibitem{zhou-isita18}
F.~Zhou and B.~M.~Kurkoski, ``Shaping gain of lattices
  based on convolutional codes and {Construction~A},'' in
  \emph{Proc.~2018 International Symposium on Information Theory and its Applications}, 2018, pp.~183--187.
\BIBentrySTDinterwordspacing

\bibitem{costa-2017}
S.~I.~R.~Costa, F.~Oggier, A.~Campello, J.-C.~Belfiore, and E.~Viterbo,
  \emph{Lattices Applied to Coding for Reliable and Secure Communications}, ser. {SpringerBriefs} in {Mathematics}.\hskip 1em plus 0.5em minus
  0.4em\relax Cham, Switzerland: Springer International Publishing, 2017.
\BIBentrySTDinterwordspacing

\bibitem{Lin-2004}
S.~Lin and D.~J.~Costello, \emph{Error Control Coding}, 2nd~ed.\hskip 1em plus 0.5em minus 0.4em\relax Upper Saddle River, NJ, USA: Prentice-Hall, Inc., 2004.
\BIBentrySTDinterwordspacing

\bibitem{Shao-com03}
R.~Y.~Shao, S.~Lin, and M.~P.~C.~Fossorier, ``Two decoding algorithms for tailbiting codes,'' \emph{IEEE Transactions on Communications}, vol.~51, no.~10, pp.~1658--1665, Oct.~2003.
\BIBentrySTDinterwordspacing

\bibitem{Wang-pimrc96}
Y.-P.~E.~Wang and R.~Ramesh, ``To bite or not to bite---a study of tail bits versus tail-biting,'' in \emph{Proc.~7th IEEE International Symposium on Personal, Indoor and Mobile Radio Communications}, vol.~2, 1996, pp.~317--321.
\BIBentrySTDinterwordspacing

\bibitem{rosnes-it12}
E.~Rosnes, {\O}.~Ytrehus, M.~A.~Ambroze, and M.~Tomlinson,
  ``Addendum to “{A}n efficient algorithm to find all small-size stopping sets of low-density parity-check matrices”,'' \emph{IEEE Transactions on Information Theory}, vol.~58, no.~1, pp.~164--171, Jan.~2012.
\BIBentrySTDinterwordspacing

\bibitem{Zhou-isit21}
F.~Zhou, A.~Fitri, K.~Anwar, and B.~M.~Kurkoski, ``Encoding and decoding
  {Construction~D'} lattices for power-constrained communications,'' in \emph{Proc.~2021 IEEE International Symposium on
  Information Theory}, 2021, pp.~1005--1010..
\BIBentrySTDinterwordspacing

\bibitem{Conway-it82}
J.~H.~Conway and N.~J.~A.~Sloane, ``Fast quantizing and decoding and algorithms for lattice quantizers and codes,'' \emph{IEEE Transactions on Information Theory}, vol.~28, no.~2, pp.~227--232, Mar.~1982.
\BIBentrySTDinterwordspacing

\bibitem{Viterbo-it99}
E.~Viterbo and J.~Bouros, ``A universal lattice code decoder for fading
  channels,'' \emph{IEEE Transactions on Information Theory}, vol.~45, no.~5, pp.~1639--1642, Jul.~1999.
\BIBentrySTDinterwordspacing

\end{thebibliography}
